\documentclass[iop, apj, numberedappendix, twocolappendix]{emulateapj}

\usepackage{natbib}  % added by simon apr-2015
\usepackage{color}
\usepackage{epsfig,graphicx,graphics}

\shorttitle{Dust trapping in HD~142527}
\shortauthors{Casassus et al.}

\begin{document}


\title{A compact concentration of large grains in the HD~142527 protoplanetary dust trap} 


\author{Simon Casassus\altaffilmark{1,2},
  Chris Wright\altaffilmark{3},
  Sebastian Marino\altaffilmark{1,2},
  Sarah T. Maddison\altaffilmark{4},
  Al Wootten\altaffilmark{5},
  Pablo Roman\altaffilmark{6,2},
  Sebastian P\'erez\altaffilmark{1,2},
  Paola Pinilla\altaffilmark{7},
  Mark Wyatt\altaffilmark{8},
  Victor Moral\altaffilmark{6,2},
  Francois M\'enard\altaffilmark{9},
  Valentin Christiaens\altaffilmark{1,2},
  Lucas Cieza\altaffilmark{10},
  Gerrit van der Plas\altaffilmark{1,2} 
}

%  and Ivan R. King\altaffilmark{1}}
%\affil{Astronomy Department, University of California,
%    Berkeley, CA 94720}
%
%\author{C. D. Biemesderfer\altaffilmark{4,5}}
%\affil{National Optical Astronomy Observatories, Tucson, AZ 85719}
%\email{aastex-help@aas.org}
%
%\and
%
%\author{R. J. Hanisch\altaffilmark{5}}
%\affil{Space Telescope Science Institute, Baltimore, MD 21218}


\altaffiltext{1}{Departamento de Astronom\'{\i}a, Universidad de Chile, Casilla 36-D, Santiago, Chile}
\altaffiltext{2}{Millennium Nucleus ``Protoplanetary Disks'', Chile}
\altaffiltext{3}{School of Physical, Environmental and Mathematical Sciences, UNSW Canberra, PO Box 7916, Canberra BC 2610, Australia}
\altaffiltext{4}{Centre for Astrophysics \& Supercomputing, Swinburne University of Technology, PO Box 218, Hawthorn, VIC 3122, Australia}
\altaffiltext{5}{National Radio Astronomy Observatory, 520 Edgemont Road, Charlottesville, VA 22903-2475, USA}
\altaffiltext{6}{Center for Mathematical Modeling, Universidad de Chile,  Av. Blanco Encalada 2120 Piso 7, Santiago, Chile}
\altaffiltext{7}{Leiden Observatory, Leiden University, P.O. Box 9513, 2300RA Leiden, The Netherlands}
\altaffiltext{8}{Institute of Astronomy, University of Cambridge, Madingley Road, Cambridge CB3 0HA, United Kingdom}
\altaffiltext{9}{UMI-FCA, CNRS / INSU France (UMI 3386), at Departamento de Astronom\'{\i}a, Universidad de Chile, Santiago, Chile}
\altaffiltext{10}{Facultad de Ingeniería,  Universidad Diego Portales, Av. Ejército 441, Santiago, Chile}


\begin{abstract}

A pathway to the formation of planetesimals, and eventually giant
planets, may occur in concentrations of dust grains trapped in
pressure maxima. Dramatic crescent-shaped dust concentrations have
been seen in recent radio images at sub-mm wavelengths. These disk
asymmetries could represent the initial phases of planet formation in
the dust trap scenario, provided that grain sizes are spatially
segregated. A testable prediction of azimuthal dust trapping is that
progressively larger grains should be more sharply confined and
furthermore the trapped grains should follow a distribution that is
markedly different from the gas. However, gas tracers such as CO and
the infrared emission from small grains are both very optically thick
where the submm continuum originates, so observations have been unable
to test the trapping predictions or to identify compact concentrations
of larger grains required for planet formation by core-accretion. Here
we report multifrequency observations of HD~142527, from 34~GHz to
700~GHz, that reveal a compact concentration of $\sim$cm-sized grains,
with a few Earth masses, embedded in a large-scale crescent of
$\sim$mm-sized particles. The emission peaks at wavelengths shorter
than $\sim$1~mm are optically thick and trace the temperature
structure resulting from shadows cast by the inner regions. Given this
temperature structure, we infer that the largest dust grains are
concentrated in the 34 GHz clump. We conclude that dust trapping is
efficient for approximately cm-sized grains and leads to enhanced
concentrations, while the smaller grains largely reflect the gas
distribution.

\end{abstract}



\section{Introduction}\label{sec:intro}

Giant planet formation occurs in the first few million years following
stellar birth, while the parent protoplanetary disk is still gas-rich
\citep[][]{1995Natur.373..494Z}. However, the classical debate on the
formation mechanism, if envelope accretion onto a rocky core,
i.e. core-accretion \citep[][]{Pollack1996Icar..124...62P}, or
gravitational instability \citep[][]{Kuiper1951PNAS...37....1K}, is
stalled without observations.  The pathway to giant planet formation
determines the initial configuration and compositions of the product
planetary systems. Current trends in theory contemplate a variety of
formation scenarios, notably second-generation core-accretion at large
radii following the formation of a giant closer-in
\citep[][]{Ayliffe2012MNRAS.423.1450A, Sandor2011ApJ...728L...9S}.
Models predict that young giant protoplanets carve a deep gap in the
dust component of protoplanetary disks, and a shallower gap in the gas
\citep[][]{PaardekooperMellema2006, Fouchet2010}.  The clearing of the
protoplanetary gap is thought to underlie the class of `transition'
disks. A local pressure maximum develops at the outer edge of the gap,
that can trap and pile-up the larger grains
\citep[][]{Zhu_Stone_2014ApJ...795...53Z}, which would otherwise
rapidly migrate inwards due to aerodynamic drag
\citep[][]{Weidenschilling1977MNRAS.180...57W} (as in the so-called
`meter-size barrier', which at 50--100~AU corresponds to mm-sized
particles).  Local pressure maxima could also occur in lopsided disks
with a stellar offset \citep[][]{Mittal2015ApJ...798L..25M}. Whichever
is origin, the development of a local pressure maximum may promote the
pile-up and growth of dust grains, and the formation of large dust
clumps that could eventually collapse in planetary cores
\citep[][]{Ayliffe2012MNRAS.423.1450A, Lyra2009A&A...497..869L,
  Sandor2011ApJ...728L...9S}.

Recent observational progress brought by the Atacama Large Millimeter
Array (ALMA) fits in the scenario of second-generation
planet-formation at large stellocentric radii. The observational
identification of so-called dust traps, in the form of concentrations
of mm-sized dust \citep[][]{Casassus2013Natur,
  vanderMarel2013Sci...340.1199V} in the outer regions of transition
disks, suggests that azimuthal dust trapping may occur. An important
prediction of this scenario is that progressively larger grains should
be more sharply confined \citep[][]{Birnstiel2013A&A...550L...8B,
  LyraLin2013ApJ...775...17L}, possibly leading to the formation of
boulders and planetesimals. Another consequence is that the larger
dust grains should have a distribution that is markedly different from
the gas. Observed contrast ratios of 30 to 100 between extrema in the
sub-mm continuumm (at coarse angular resolutions) would seem too large
to reflect an equally lopsided gas distribution. However, a
theoretical question remains as to the impact of enhanced cooling in
the dust trap on the physical conditions. In observational terms, the
use of gas tracers such as CO or the infrared (IR) emission from small grains are
both very optically thick where the ALMA thermal continuum originates,
so that the previous observations \citep[][]{Casassus2013Natur,
  vanderMarel2013Sci...340.1199V,Perez_L_2014ApJ...783L..13P} could
not test the trapping predictions. For instance, in IRS~48 the outer
ring is optically thick in the mid-IR emission \citep[][their
  Sec.~4.1]{Bruderer_2014A&A...562A..26B}, as is CO(6-5) whose
azimuthal structure is largely determined by foreground absorption.

Since emission at a given wavelength is dominated by grains of a
matching size, multi-wavelength radio observations can confirm if the
crescent-shaped continua seen in transition disks are indeed due to
the dust trap phenomenon. In this work we compare new ALMA
high-frequency radio continuum observations of HD~142527, obtained at
700~GHz (ALMA band~9; wavelengths of $\sim$0.43~mm, described in
Appendix, Sec.~\ref{sec:ALMAobs}), with resolved images at 34~GHz
(8.8~mm) acquired at the Australia Telescope Compact Array (ATCA, see
Sec.~\ref{sec:ATCAobs}). In Section~\ref{sec:ALMAresults} we show that
the optical depths are of order $\sim$1 at 345~GHz, so that the sub-mm
spectral trends observed with ALMA correspond to optical depths
effects rather than to dust trapping. In Sec.~\ref{sec:ATCAresults} we
explain that the ATCA observations reveal a compact clump at 34~GHz
embedded in the sub-mm crescent, which is reflected by an opacity
spectral index corresponding to larger grains, as expected in the dust
trapping scenario. In Sec.~\ref{sec:models} we discuss the
observational results in terms of radiative transfer predictions based
on a parametric dust trap mode.  Sec.~\ref{sec:conclusions} summarises
our results.

\section{An optically thick sub-mm continuum}  \label{sec:ALMAresults}

\begin{figure}
\begin{center}
  \includegraphics[width=0.8\columnwidth,height=!]{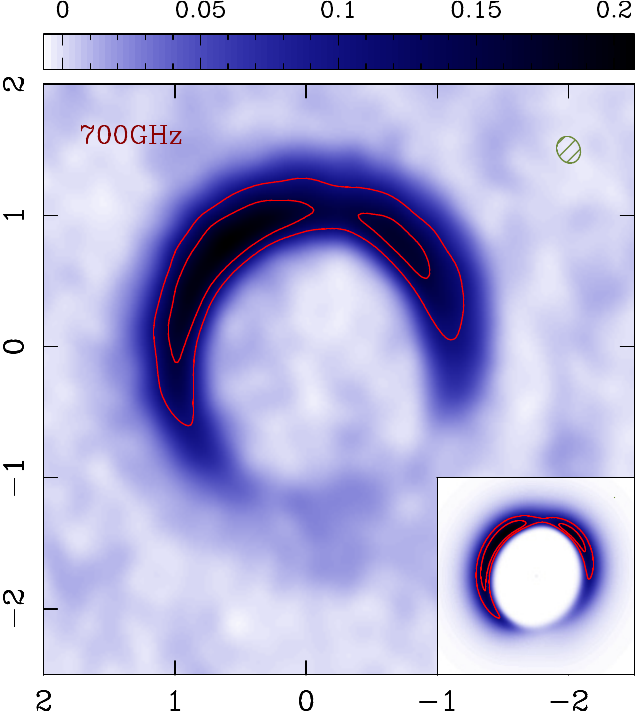}
\end{center}
\caption{\small ALMA band~9 observations of HD~142527, and comparison
  with synthetic predictions.  $x-$ and $y-$ axes indicate angular
  offset in arcsec along right-ascension (RA) and declination (Dec)
  relative to the stellar position, at the origin of coordinates. The
  color scale shows a restored image of the ALMA band~9 at 700~GHz,
  with contours at 0.5 and 0.75 times the peak intensity. The wedge
  indicates specific intensity in Jy~beam$^{-1}$, with a beam of
  0.21$\times$0.17~arcsec (the beam ellipse is shown on the upper
  right corner). The inset shows emergent intensities predicted from
  the dust trap model (see Sec.~\ref{sec:models}), including the impact on grain temperatures due
  to shadowing from a tilted inner disk. Contours are at 0.5 and 0.75
  times the peak, no smoothing has been applied. The side of the inset
  corresponds to 3.5~arcsec.
  \label{fig:b9} }
\end{figure}

The ALMA band~9 continuum (Figure~\ref{fig:b9}) has a morphology that
is intermediate between the broken ring seen in the thermal IR at
18~$\mu$m \citep[][]{Fujiwara2006,Verhoeff2011A&A...528A..91V} and the
345~GHz crescent seen in ALMA band~7
\citep[][]{Casassus2013Natur}. The maximum radial width of the
crescent, measured at half-maximum, is about 0.6\arcsec, so is
resolved with a beam size of 0.25\arcsec.

A traditional means to quantify the spectral variations as a function
of position is the spectral index map $\alpha(\vec{x})$, with a
power-law parameterisation of the specific intensity $I_\nu =
I_{\nu\circ}(\nu / \nu_\circ)^{\alpha(\vec{x})}$. We find that the
spectral index between 345~GHz and 700~GHz, $\alpha_{345}^{700}$, as
well as the slope of the spectrum across the 0.13~GHz frequency lever
within ALMA band~7, $\alpha_{345}$, both anti-correlate with the
continuum crescent (see Sec.~\ref{sec:alpha}), meaning that the
azimuthal extension is greater at shorter wavelength. However, while
the band~7 and band~9 continua peak between $10$h and $11$h as a
function of azimuth along the outer ring, the spectral index maps
exhibit a common minimum at $\sim$1h, i.e. at the northern ansa of the
ring. This coincidence is indicative of a projection effect, as in
limb-brightening: if column densities are greater at 1h, the increased
optical depths will result in flatter spectra.

%this requires to align and bring the
%multi-frequency visibilities to a common $uv$-plane coverage,

%(this corresponds to the location where the intersection of the
%outer disk plane with the plane of the sky meets the outer disk)

%\footnote{the contrast ratio between the specific intensity extrema in
%  the outer disk is $>$20 at 700~GHz, and 28 at
%  345~GHz \citep[][]{Casassus2013Natur}, in line with the dust trap
%  predictions \citep[][]{Birnstiel2013A&A...550L...8B,vanderMarel2013Sci...340.1199V}. However,
%  this ratio is not a good proxy of morphology since the low-level
%  signal is structured, i.e. it is more complex than a smooth
%  fundamental-harmonic modulation with azimuth, and is also affected
%  by the underlying image synthesis systematics, derived in particular
%  from the need for self-calibration}.
%

%the four intra-band central frequencies of the ALMA data can also be
%used to build an intra-band~7 spectral index map

The spectral variations inferred from the multi frequency
morphological trends can be cast into sky maps for the optical depth
$\tau(\vec{x}) = \tau_\circ(\vec{x}) \times
(\nu/\nu_\circ)^{\beta_s(\vec{x})}$ and line-of-sight temperature
$T_s(\vec{x})$, as a function of angular position $\vec{x}$. We fit
the observations with the intensities emergent from a uniform slab
(hereafter `grey-body', see Sec.~\ref{sec:ALMAalign} and Sec.~\ref{sec:Ttaubet}),
\[
I_\nu^m(\vec{x}) = B_\nu(T_s(\vec{x}))\left[1 - \exp( -\tau(\vec{x}) )
  \right].\] The inferred temperature field approximates the
opacity-weighted average temperature along the line of sight (see
Sec.~\ref{sec:biases} for a discussion of biases).  In synthesis, as
explained in Sec.~\ref{sec:Ttaubet}) and illustrated in
Fig.~\ref{fig:Ttaubet_b9}, the crescent reaches optical depths close
to 1 at 345~GHz, even in a rather coarse beam (which dilutes the
signal).  The optical depth maximum is coincident with the location
where the sub-mm spectral index ($\alpha_{345}^{700}$ and
$\alpha_{345}$) is minimum, thus in agreement with an interpretation
of the spectral index variations in terms of optical depth effects,
rather than genuine variations of the underlying dust grain
populations. Another consequence of the high submm optical depths is
that the double-peaked crescent morphology, best seen in band~9,
reflects structure in the temperature field rather than in the density
field inside the dust trap. The local band~9 emission maxima, at 10.5h
and 1.5h, are separated by a minimum in temperature at 0.5h.

%Interestingly, the crescent-shaped sub-mm continuum coincides with a
%similarly shaped molecular line decrement in CO(3-2)
%(Fig.~\ref{fig:decrements}). This coincidence is consistent with an
%optically thick 345~GHz continuum: given the very high opacities of CO
%line emission, the crescent-shaped continuum could be leveled out by
%the more extended and uniform disk emission in CO(3-2).
%

\section{ATCA/ALMA multi-frequency results}  \label{sec:ATCAresults}

\subsection{ A 34~GHz clump embedded in the sub-mm crescent}

\begin{figure*}
\begin{center}
\includegraphics[width=0.7\textwidth,height=!]{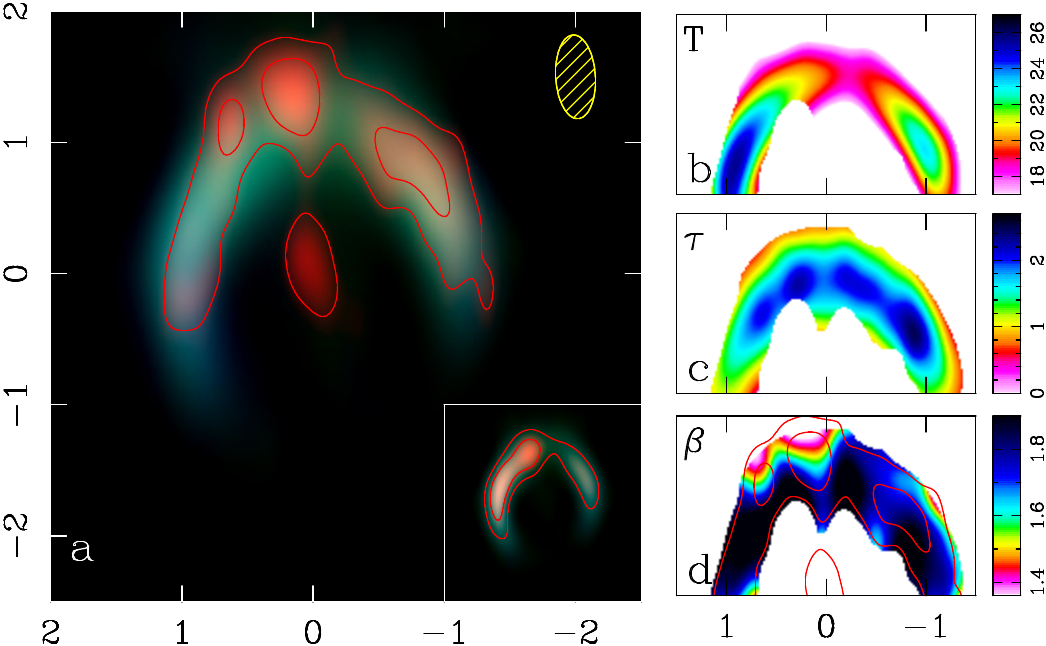}
\end{center}
\caption{\small Multi-frequency continuum highlighting larger grains
  at 11.5h. The data have been filtered for the ATCA response.  {\bf
    a}: RGB image with the ATCA image in red (with corresponding
  contours at 0.4 and 0.7 times the peak intensity at
  147$\mu$Jy~beam$^{-1}$), ALMA band~7 in green, and ALMA band~9 in
  blue (see Fig.~\ref{fig:MCtests} for a panel with the same shown
  images separately).  The inset shows emergent intensities predicted
  from the dust trap model, after filtering for the ATCA response,
  with 34~GHz in red contours at 0.5 and 0.75 times the peak. The side
  of the inset corresponds to 3.5~arcsec.  {\bf b}: Observed grey-body
  temperature map. {\bf c}: optical depths at 345~GHz. {\bf d}:
  emissivity index map $\beta_s(\vec{x})$, with ATCA contours in red.
  The $\beta_s$ minimum at 11.5h indicates the presence of larger
  grains.  \label{fig:ATCAref_RGB} }
\end{figure*}

Resolved ATCA observations of HD~142527 in the optically thin 34~GHz
continuum reveal structure inside the sub-mm-opaque crescent, as shown
in Fig.~\ref{fig:ATCAref_RGB} (see also
Fig.~\ref{fig:ATCAselfcal}). The lower levels at 34~GHz follow the
Rayleigh-Jeans extension of the hotter dust seen in ALMA
band~9. However, the peak signal stems from clumpy emission at
$\sim$11.5h and a longer arc at 1h to 2h. Both regions coincide fairly
closely with molecular decrements (Sec.~\ref{sec:decrements}).  Given
the thermal noise in the field, the 11.5~h clump is significant at
10$\sigma$, and it is robust against $uv-$filtering biases.  To test
for interferometer filtering artefacts, we ran Monte-Carlo simulations
of ATCA observations on a deconvolved model image of the band~7 data
(see Sec.~\ref{sec:MCtests}), thus bringing the multi-frequency data
to an approximately common Fourier basis. Fig.~\ref{fig:ATCAref_RGB}
shows that the ATCA clump stands out compared to the filtered ALMA
maps.

%This offset is not significant enough to warrant modeling, but it is
%nonetheless intriguing that the 11.5h ATCA peak matches one of the
%local minima in HCO$^+$.

The peak 34~GHz intensity at 11.5h reaches 147$\mu$Jy~beam$^{-1}$,
with a $0.64\arcsec\times0.30$\arcsec~beam, and has an equivalent
brightness temperature of only 0.5~K, hence the signal is optically
thin (or else it is beam-diluted). For the fiducial density of water,
and for a single grain size of 1~cm with a circular cross section, at
20~K, we find that the observed flux density from the unresolved 11.5h
clump corresponds to $>$7.4~M$_\oplus$ of dust. The inequality
accounts for the possibility of locking more mass in grains of
different sizes that do not emit appreciably at 34~GHz.

%For comparison, if the total clump mass, including gas, is
%$\sim$1~M$_\mathrm{jup}$, then the Hill radius is $\sim$8~AU and
%smaller than the ATCA beam, so a gravitationally could be
%gravitationally bound. \textcolor{red}{***Replace with an estimate of
%  Q for the dust population. Is this worth stating?***}.

A compact signal coincident with the stellar position is conspicuous
in the ATCA map. Its spectral index at ATCA frequencies is
$\alpha_{20}^{45} = 1.0 \pm 0.2$, which extrapolates to the level of
the faint circumstellar signal seen in ALMA band~7
\citep[][]{Casassus2013Natur,2013PASJ...65L..14F}. These spectral
index values are consistent with free-free emission associated with a
stellar wind, or with stellar accretion.

\subsection{Evidence for dust trapping}  \label{sec:Ttaubet_main}

In another application of the grey-body diagnostics, we computed the
optical depth and opacity index ($\beta_S$) maps corresponding to the
ATCA and ALMA datasets shown in Fig.~\ref{fig:ATCAref_RGB}a.  The
resulting optical depth map shown in Fig.~\ref{fig:ATCAref_RGB}c
confirms the levels at 345~GHz inferred from the ALMA band~7 and
band~9 comparison.  The ATCA clump at 11.5h translates to a minimum in
$\beta_S(\vec{x})$ (Fig.~\ref{fig:ATCAref_RGB}d). This is an
indication that dust grains in this clump are larger than in the rest
of the crescent.  As summarised in Fig.~\ref{fig:Ttaubet}, the
variations in $\beta_S$ are significant at $\sim$7~$\sigma$.

%Hence CO does not stand out over
%the optically thick continuum.

%(or
%perhaps also due to gravity from compact objects)

%\textcolor{red}{*** add a few words about
%  the HCO+ production - cannot be UV radiation in the outer disk since
%  we do not see the warp shadows in HCO+ ***}
%

%\textcolor{red}{******
%  quantify more precisely***} compared to the actual optical depth in
%the synthetic dust trap, the grey-body temperature is close to the
%average temperature of grains weighted by their opacities along the
%line of sigh.t
%

\section{Predictions based on a synthetic dust trap model}  \label{sec:models}

\subsection{Biases of the grey-body diagnostics} \label{sec:biases}

The grey-body line-of-sight diagnostic is based on 2-dimensional sky
maps for $\beta_S(\vec{x})$ and $T_S(\vec{x})$, while the actual
physical conditions are 3-dimensional.  Since the crescent is
optically thick in band~9, the temperature is mainly set by the higher
frequencies, and probably overestimates the temperature in regions
closer to the midplane.  Hence the actual optical depths could be
higher than inferred from grey-body diagnostics. In order to assess
biases and test the dust trap interpretation, we applied the same
observational diagnostic, based on the grey-body fits, to synthetic
radiative transfer predictions of a parametrised dust trap model meant
to approximate the phenomenon observed in HD~142527.  The synthetic
model was implemented in RADMC3D \citep[][]{RADMC3D0.39} (Methods), and
centers the dust trap at 11.5h. At the native resolution of the
radiative transfer, the grey-body optical depth map reproduces the
input optical depths very closely. However, both optical depth and
temperatures are biased downwards by $\sim$50\% after filtering by the
ATCA response.  The location of the minimum in the grey-body opacity
index, $\beta_S(\vec{x})$, coincides with the input dust trap, which
is fairly optically thick in band~7. However the optical depths in the
synthetic trap are not high enough to result in the observed deep
molecular decrements - their properties could be reproduced by
prescribing gas temperatures cooler than the dust
(Sec.~\ref{sec:decrements}). In particular the existence of decrements
in CO(2-1)  \citep[][]{Perez2015ApJ...798...85P}, so at frequencies with
lower optical depths, suggests that there is ample room for other
effects.  The decrements may also correspond to structural features of
the disk, i.e. shadowing in lower scale heights under the 34~GHz
clumps, due to midplanes cooler than in the rest of the outer disk at
the same stellocentric radius.  A mechanism for such a local
`pinching' of the outer disk may be found in the dust trap
scenario. In radiative equilibrium the larger grains tend to be
cooler, so that hydrostatic scale heights will be lower under the dust
trap. The shadowed CO is cooler, or possibly frozen out and condensed
onto grain surfaces.

\subsection{Temperature decrement due to shadowing on the outer-ring}

The temperature structure inferred from the band~9 observations are
readily explained by the model, which includes a tilted inner disk
within 20~AU, as implied by the shadows in scattered light cast on the
outer disk \citep[][]{Marino2015ApJ...798L..44M}. Fig.~\ref{fig:b9}
shows that the predicted emergent intensities in band~9 are
double-peaked, much like the observations. The warped structure
results in shadows projected on the outer disk where the stellar
heating is blocked by the inner disk at PAs of 0.5h and 6.5h. The
similarity with the observations (Fig.~\ref{fig:Ttaubet_b9}) is
remarkable considering the idealizations of the synthetic disk, which
assumes a circular cavity, power-laws for the grain size
distributions, and steady-state dust trapping prescriptions. Another
idealisation of the model is steady state passive heating.

The cooling timescale is $\tau_C = U / (\sigma_B T^4)$, where $U$ is
the internal energy per unit area, and $\sigma_B$ is the
Stefan-Boltzmann costant. The condition that $\tau_c$ must be shorter
than the crossing time under shadows that cover $\sim$20\degr at
140~AU \citep[][ and
  Fig.~\ref{fig:warpshadows}]{Marino2015ApJ...798L..44M}, is satisfied
for surface densities lower than 50~g~cm$^{-2}$, which is indeed the
maximum surface density in the model. It is thus reasonable to expect
that the double-peaked structure of the continuum emission is
determined by the temperature field, and that the mm-grains are
otherwise uniformly spread along the crescent. The modulation in
temperature along the crescent, caused by the inner warp, 
impacts on the thermodynamics of the disk. This modulation imparts a
periodic forcing to the outer disk, whose dynamical consequences
should be investigated, and that could perhaps explain the observed
spiral pattern.

\subsection{Very lopsided disks?}

The spectral trends in HD~142527 at ALMA frequencies can thus be
accounted for with optical depth effects, such that no trapping is
required for grains up to $1~$mm in size. In other words, the large
sub-mm crescent \citep[][]{Casassus2013Natur} mostly reflects the gas
background, with relatively inefficient trapping, so that the observed
contrast ratio of $\sim$30 is accounted for with a contrast of
20 in the gas.  Marked gaseous asymmetries with contrasts of order 10
are seen in MHD simulations of dynamical clearing by planet
formation \citep[][]{Zhu_Stone_2014ApJ...795...53Z}, and could perhaps reach
higher values in models incorporating dead-zones with discontinuous
turbulence prescriptions \citep[][]{Regaly2012MNRAS.419.1701R}, or with
enhanced cooling inside the dust trap via dust-gas coupling. Thus we propose
that other sub-mm asymmetries could partly reflect the gas, perhaps
even with contrasts of order $\sim$100 such as in IRS~48, and most
likely in the more moderate contrasts, such as in HD~135344B and
SR~21 \citep[][]{Perez_L_2014ApJ...783L..13P}.

\section{Conclusion} \label{sec:conclusions}

In summary, the multi-frequency observations of HD~142527 are
consistent with the dust trap scenario. New ALMA observations have
shown that the mm-grains are so abundant in the crescent that efforts
to observe grain growth at sub-mm wavelength are thwarted by optical
thickness. Optical depth effects account for the spectral trends at
ALMA frequencies, while shadowing of UV-optical radiation by the
central warp explains the temperature structure of the crescent.
However, at transparent cm-wavelengths a compact clump of
approximately cm-sized grains is embedded within the broader crescent
of smaller mm-grains (the exact grain sizes depends on the size
distributions).  These phenomena are expected in the context of the
aerodynamic coupling of dust grains with pressure gradients, given the
physical conditions in the passively heated outer disk of
HD~142527. Prospects are good for witnessing planetesimal formation by
dust trapping with radio observations of HD~142527 at cm-wavelength.

\acknowledgments

We thank Cornelis Dullemond for interesting discussions and help with
RADMC3D. Based on observations acquired at the ALMA Observatory,
through programs {\tt \small \sc ALMA\#2011.0.00465.S} and {\tt \small
  ALMA\#2011.0.00318.S}. ALMA is a partnership of ESO, NSF, NINS, NRC,
NSC, ASIAA. The Joint ALMA Observatory is operated by ESO, AUI/NRAO
and NAOJ. The Australia Telescope Compact Array is part of the
Australia Telescope National Facility which is funded by the
Commonwealth of Australia for operation as a National Facility managed
by CSIRO. Financial support was provided by Millennium Nucleus
RC130007 (Chilean Ministry of Economy), and additionally by FONDECYT
grant 1130949. CMW acknowledges support from ARC Future Fellowship
FT100100495.  S.M. acknowledges CONICYT-PCHA / Magister Nacional /
2014-22140628.  PR and VM acknowledge support from ALMA-CONICYT grant
31120006. This work was partially supported by the Chilean
supercomputing infrastructure of the NLHPC (ECM-02).

\appendix

\section{Observations}  \label{sec:obs}

HD~142527, at $\sim$140~pc, is a promising laboratory to observe
on-going gas giant planet formation, and confirm the existence of the
dust trap phenomenon. This transition disk has a close to face-on
orientation.  Infrared observations found an inner disk,
$\sim$10~AU \citep[][]{Fujiwara2006} in radius, and surrounded by a
particularly large gap \citep[][]{Fukagawa2006}. A disrupted outer
disk \citep[][]{Casassus2012ApJ...754L..31C} beyond 140~AU is indicative of
dynamical clearing of the gap by giant planet formation. The cavity is
devoid of bodies larger than
$\sim$4~M$_\mathrm{jup}$ \citep[][]{Casassus2013Natur}, but an accreting
companion has been reported at $\sim$10~AU \citep[][]{Biller2012,
  Close2014ApJ...781L..30C, Rodigas_2014ApJ...791L..37R} (with a
binary mass ratio of 0.01 to 0.1). The bulk of the mass lies in the outer
disk, whose crescent
morphology \citep[][]{2008Ap&SS.313..101O,Casassus2013Natur} approximately
anti-correlates with molecular-line emission \citep[][]{Casassus2013Natur,
  vdP_2014ApJ...792L..25V, Perez2015ApJ...798...85P}. A complex array
of trailing spiral arms sprouts away from the outer disk
 \citep[][]{Fukagawa2006, Casassus2012ApJ...754L..31C,
  Rameau2012A&A...546A..24R, Canovas2013A&A...556A.123C,
  Avenhaus2014ApJ...781...87A, Christiaens2014ApJ...785L..12C}.

In this Section we provide details on the datasets that sustain our
multi-frequency analysis. Sec.~\ref{sec:ALMAobs} presents the new ALMA
observations, Sec.~\ref{sec:ATCAobs} presents the new ATCA data and
Sec.~\ref{sec:ALMAaux} summarizes previously published ALMA
data. Sec.~\ref{sec:imagesynthesis} briefly describes our image
synthesis procedure, while Sec.~\ref{sec:ALMAalign} and
Sec.~\ref{sec:alpha} describe the alignment of the multi-frequency
ALMA data and associated spectral index trends and
Sec.~\ref{sec:ATCAALMAalign} describes the alignment of the ATCA and
ALMA datasets.

\subsection{ALMA band~9 observations} \label{sec:ALMAobs}

ALMA Band 9 observations of HD~142527 for program {\tt \small \sc
  JAO.ALMA\#2011.0.00465.S} were carried out in 2012, in the nights of
June 3, from 02:52:53 to 04:39:10 UT, and on June 4, from 00:43:22 to
02:42:26 UT.  The precipitable water vapor (PWV) in the atmosphere was
stable between 0.54~mm and 0.62~mm on June 3, with a median value at
zenith of 0.581~mm. Conditions on June 4 were similar, with a median
PWV at zenith of 0.597~mm.
The ALMA correlator provided 1875~MHz bandwidth in four different
spectral windows at 488.28125~kHz resolution (or 211.67~m~s$^{-1}$ at
691.47308~GHz) per channel. Each spectral window was positioned in
order to target the CO(6-5) transition at 691.47308~GHz, HCO$^+$(8-7),
H$^{13}$CO$^+$(8-7), and HCN(8-7).  Only CO(6-5) was detected. This
line is the topic of a separate article.
On June 3, the measured system temperatures ranged from $\sim
1400\pm500$~K in the spectral window covering CO(6-5) to $\sim
2000\pm700$~K in the spectral window covering HCO$^+$. System
temperature values 100 to 200~K higher were recorded on June 4.  The
number of 12~m antennas available at the time of the observation was
20, although one antenna reported very large system temperatures
(DV12), and was flagged during data reduction. Excluding calibration
overheads, the total time on source for HD~142527 was 43~min on
June 3, and 49~min on June~4. 
The primary flux calibrator was Titan, while 3C279 was used as
bandpass calibrator. We used J1517-243 as phase calibrator, for which
we measure a flux density of 0.70$\pm$0.14~Jy at 702~GHz, where the
uncertainty accounts for systematic calibration errors. This
measurement is consistent with historical values for this source.
The radiometer phase correction errors were large enough to
decorrelate the signal. We used the self-calibration algorithm to
determine improved antenna-based phases that were consistent with the
continuum image of HD~142527.

\subsection{ATCA data}  \label{sec:ATCAobs}

%\paragraph{Data acquisition and calibration}

We used ATCA to observe HD~142527 in several array configurations in
May 2010 (6A), June 2010 (6C), July 2013 (6A) and August 2013 (H168),
with maximum baselines extending to 6~km. The Compact Array Broadband
Backend, or CABB \citep[][]{2011MNRAS.416..832W}, provides two
sidebands with bandwidths of 2 GHz in 2024 channels each. We centred
each sidebands at 19/21, 23/25, 33/35 and 41/43 GHz (in this work we
focus on the 33/35 dataset). Complex gains were derived from
observations of the quasar 1600-44, separated by about 2.8 degrees on
the sky from HD~142527. The gain calibrator was typically observed
every 5 to 15 min for between 1 and 3 min duration, dependent on
atmospheric conditions. Pointing checks were also made on the quasar
every 60-90 min.  The bandpass response was determined from 15 minute
observations of the quasar 1253-055 (3C 279).  The absolute flux
calibration was performed using ATCA's primary cm-band flux
calibrator, the quasar 1934-638. All subsequent data reduction and
calibration was performed with the Miriad software
\citep[][]{Miriad1995ASPC...77..433S}. The CABB setup at 33/35GHz was
used in 4 runs with different array configurations, including H168,
which allows us to estimate that the amount of flux loss in the
extended configurations is negligible. The flux densities obtained in
H168, 1.28$\pm$0.04/1.48$\pm$0.04, in Jy, for each sideband centered
at 33/35GHz, is indistinguishable from the values of
1.23$\pm$0.05/1.49$\pm$0.06, 1.34$\pm$0.11/1.48$\pm$0.11 and
1.25$\pm$0.13/1.24$\pm$0.09 obtained in 6C, 6C and 6A, respectively.

\begin{figure*}
\begin{center}
  \includegraphics[width=\textwidth,height=!]{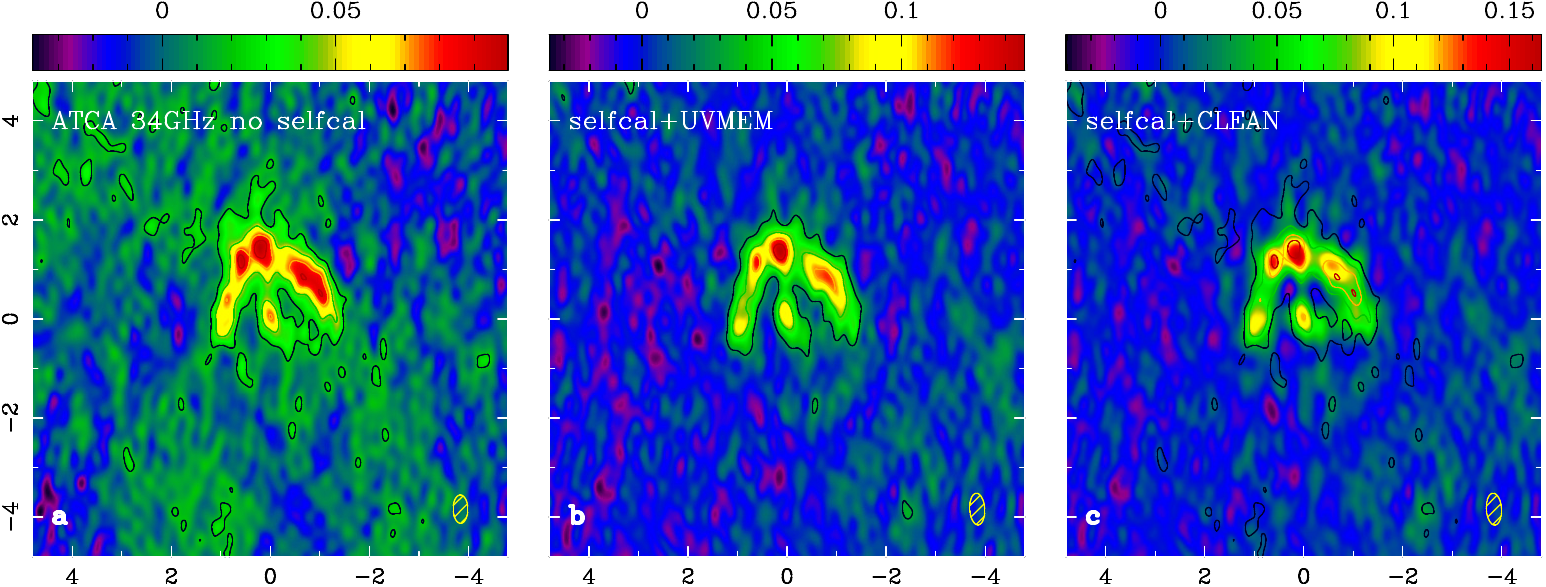}
  \end{center}
\caption{ Comparison of the restored ATCA images and the impact of
  self-calibration. The wedges indicate specific intensity in
  mJy~beam$^{-1}$.  {\bf a)} no self-calibration ( {\tt uvmem}
  restoration with a beam of $0.570\times 0.296$~arcsec$^2$).  {\bf
    b)} self-calibrated image with {\tt uvmem} restoration (with a
  beam of $0.639\times0.302$~arcsec$^2$). {\bf c)} self-calibrated
  image with {\tt CASA}  \citep[][]{CASA} {\tt clean} restoration (with a
  beam of $0.644\times0.304$~arcsec$^2$, {\tt CASA} version 4.2.2).  \label{fig:ATCAselfcal}}.
\end{figure*}

The ATCA data with the best dynamic range was obtained in Ka, i.e. at
33~GHz / 35~GHz. For efficiency we averaged each CABB 2~GHz spectral
window into 8 channels. We then brought all of the 6~km visibility
data in Ka to a common epoch, which we chose as July 2012, by
correcting for the source proper motion with a translation phase
(using the Miriad task {\tt uvedit}).  The 6~km data in Ka were then
concatenated with the task {\tt concat} from {\tt CASA}
\citep[][]{CASA} version 4.2.2. A standard clean restoration revealed
a faint modulation at low spatial frequencies in the field, atypical
of image synthesis artifacts and reminiscent of phase calibration
errors. We therefore applied selfcalibration of the visibility phases,
using the {\tt clean} model and the {\tt CASA} tasks {\tt gaincal} and
{\tt applycal}. We used a solution interval of 30~min and combined
both spectral windows for sensitivity. The {\tt clean} restoration of
the corrected visibilities results in essentially the same source
morphology as the uncorrected dataset, but with less systematics in
the field. The impact of self-calibration can be assessed by
inspection of Fig.~\ref{fig:ATCAselfcal}, where we show the restored
images obtained with our {\tt uvmem} \citep[][]{2006ApJ...639..951C,
  Casassus2013Natur} tool before and after self-calibration (more
information on this image synthesis is given in
Sec.~\ref{sec:imagesynthesis} below).

% ~/common/genfig_muppet/compar_selfcal_ATCA.pl

\subsection{Previous ALMA band~7  observations}  \label{sec:ALMAaux}

As explained in Fig.~\ref{fig:VVplot} and Sec.~\ref{sec:ALMAalign},
the calibrated dataset {\tt \small \sc ALMA\#2011.0.00465.S} is
affected by faint outliers. We therefore used the ALMA Band~7 obtained
by Fukagawa et al. \citep[][]{2013PASJ...65L..14F}, {\tt \small \sc
  \small ALMA\#2011.0.00318.S}, keeping only the last scheduling
block, observed on 10-Aug-2012. The reason for not combining the whole
dataset is because the dynamic range under the crescent is limited by
image synthesis, rather than thermal noise, and because pointing
uncertainties would degrade the beam if combined without a tedious
alignment procedure (see Sec.~\ref{sec:ALMAalign}). For instance, we
measured an offset of 0.065~arcsec between this scheduling block and
the single scheduling block in band~7 from {\tt \small \sc \small
  ALMA\#2011.0.00465.S}, comparing exclusively the common spectral
window at 342.8~GHz. These offsets are within the nominal pointing
accuracy of $\sim$0.1~arcsec. Although small, such offsets have
appreciable consequences when deriving spectral index maps.

%\textcolor{red}{*** Remove this para or else publish the companion
%  paper first - Science policy **** } In a companion paper on the
%band~9 ALMA observations \textcolor{red}{*** add ref for companion
%  paper***}, we present the CO(6-5) data included in the band~9
%observations from {\tt \small \sc \small ALMA\#2011.0.00465.S}. We
%find that the centroid of Keplerian rotation is consistent with the
%expected position of the star if the band~9 data are aligned to the
%band~7 data from {\tt \small \sc \small ALMA\#2011.0.00465.S}. Thus
%the pointing of these band~7 appear to be a safe reference.
%

In summary, we used the dataset acquired by Fukagawa et
al. \citep[][]{2013PASJ...65L..14F}, program \linebreak[4] \mbox{\tt \sc
  \small JAO.ALMA\#2011.0.00318.S} but tied the astrometric pointing
to the data previously presented by Casassus et
al. \citep[][]{Casassus2013Natur}, \linebreak[3] {\tt \small \sc
  JAO.ALMA\#2011.0.00465.S}. Further details on the astrometric
registrations are given in Sec.~\ref{sec:ALMAalign} and
\ref{sec:ATCAALMAalign}.

\subsection{Image synthesis} \label{sec:imagesynthesis}

For image synthesis we use a non-parametric least-squares modeling
technique \citep[][]{2006ApJ...639..951C} with a regularizing entropy term
(i.e. as in the family of maximum entropy methods, MEM). We call this
tool {\tt uvmem}. These deconvolved model images are `restored' by
convolution with the clean beam and by adding the dirty map of the
visibility residuals calculated using the {\tt CASA} package, for a
chosen weighting scheme. In this work we used natural weights.

The {\tt clean} algorithm is an efficient and widely accepted
technique to produce images from radio-interferometer data. To produce
consensus images that can be compared with our own {\tt uvmem} tools
we used standard {\tt CASA} {\tt clean}, i.e. Cotton-Schwabb. A
comparison between {\tt uvmem} and {\tt clean} is shown in
Fig.~\ref{fig:ATCAselfcal}.

We assume that the relative values of the visibility weights in the
ALMA data are accurate. However, the absolute values do not match the
scatter in the visibility samples recorded at each integration. In
order to extract statistics and perform least squares fits, we
therefore scaled the weights so that they correspond to 1/$\sigma^2$,
where $\sigma$ is the root mean square dispersion of all the
visibility samples recorded for a given baseline and channel. This
procedure was carried out using a tool we call {\tt uvreweight}.  A
single scale factor was applied to the weights of the whole dataset,
given by the median values of the ratios between observed scatter and
tabulated weights. Thus we respected the relative values of the
weights as provided in the calibrated dataset. We used median averages
for a robust estimate of the observed scatters, after correction by
the median absolute deviation. In this {\tt uvreweight}-processed data
the absolute value of the weights can be used to extract
statistics.

\subsection{Alignment of the multi-frequency ALMA data} \label{sec:ALMAalign}
\begin{figure}
  \begin{center}
\includegraphics[width=0.4\textwidth,height=!]{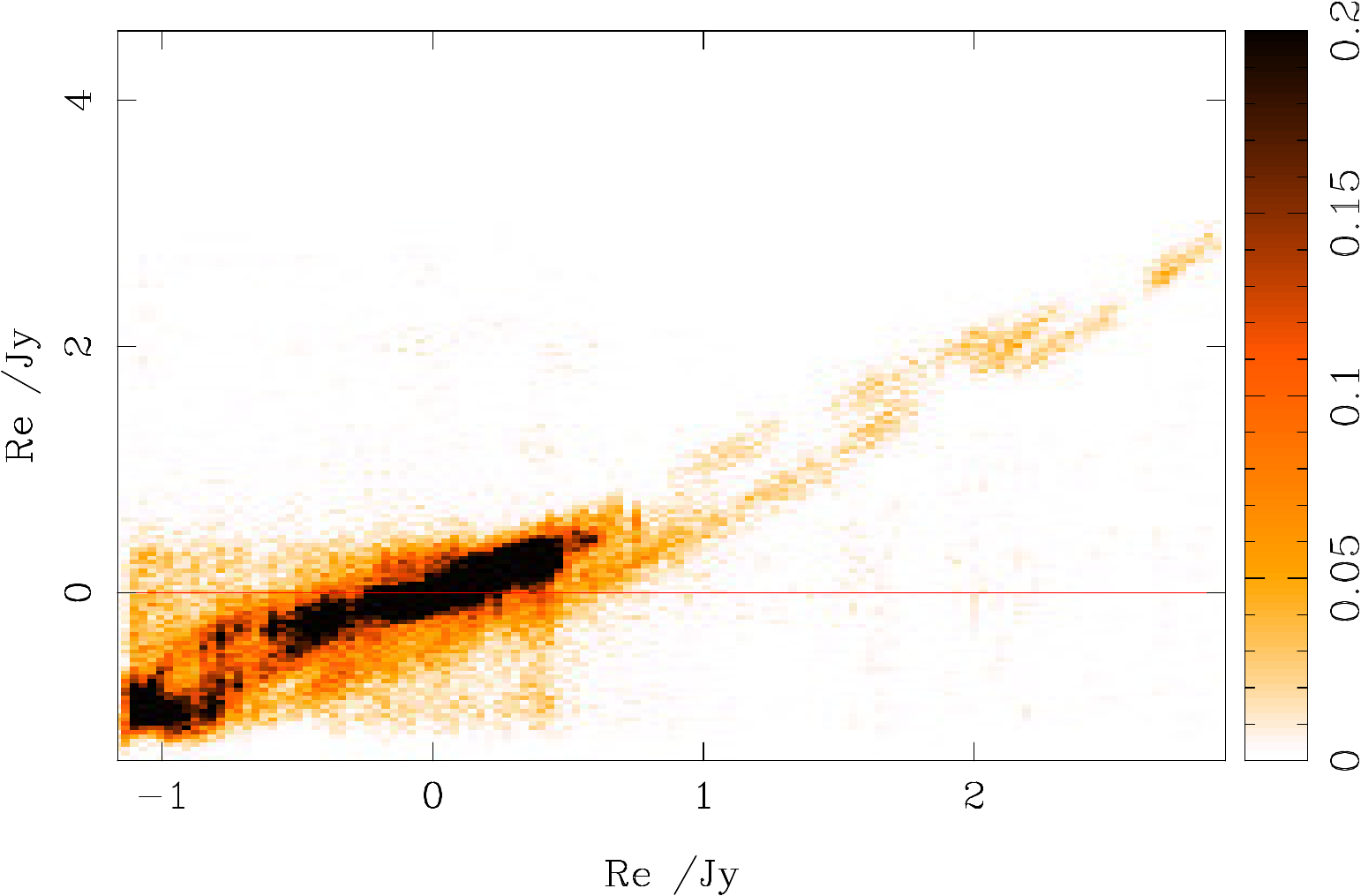}
    
\includegraphics[width=0.4\textwidth,height=!]{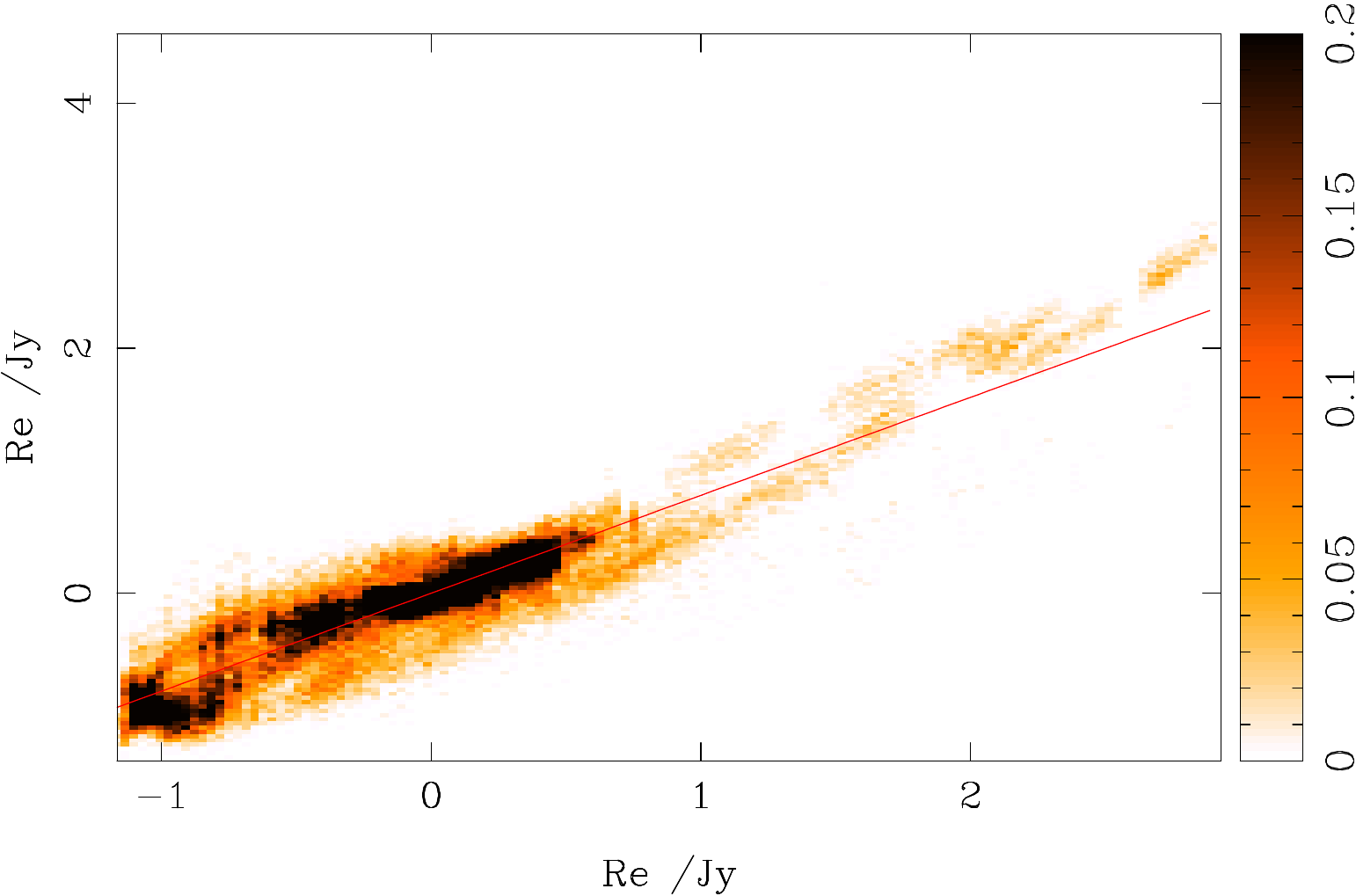}

\includegraphics[width=0.4\textwidth,height=!]{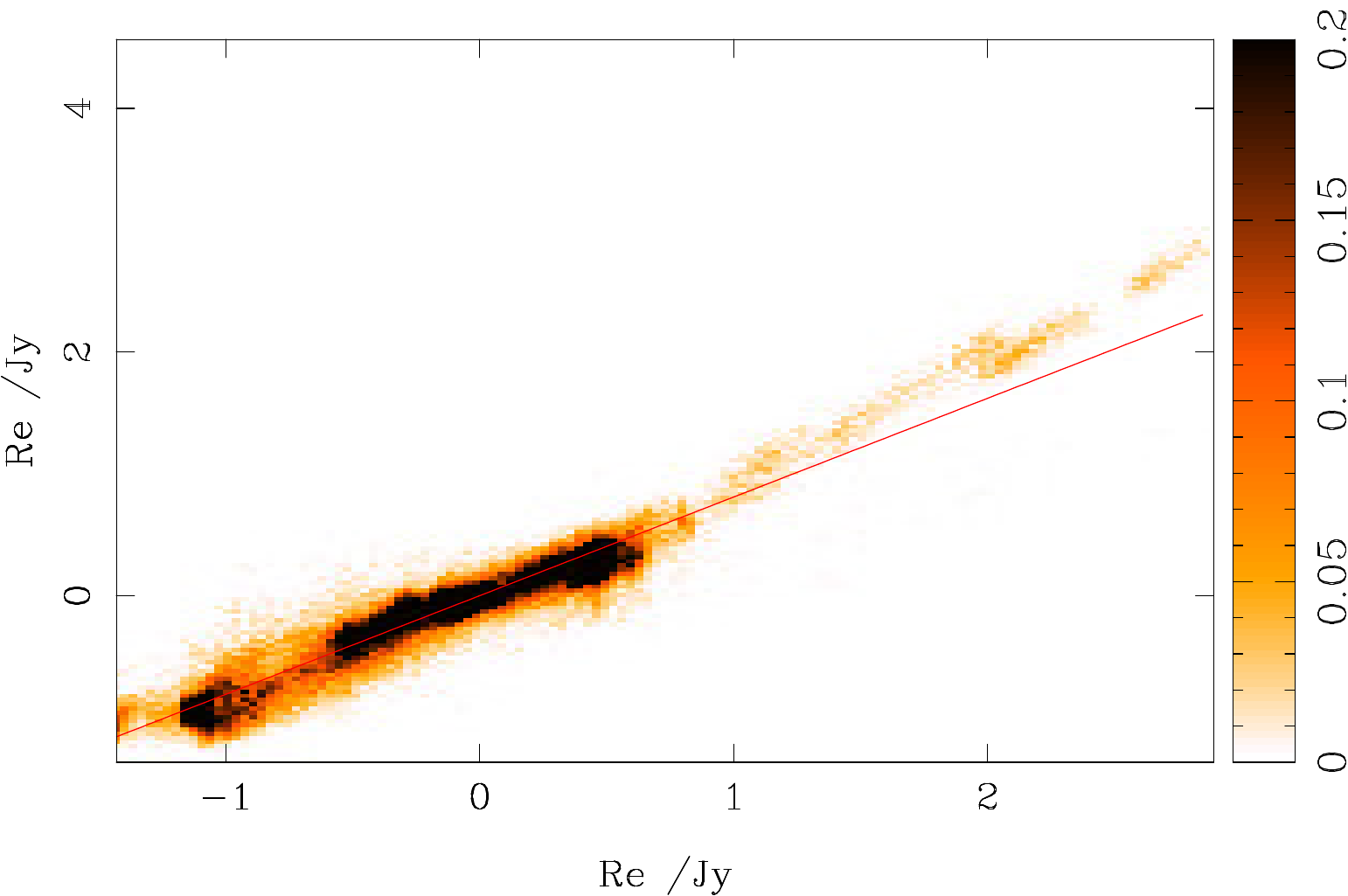}
\end{center}
  \vspace{-0.3cm}
  \caption{\small Band~7 vs. Band~9 visibility scatter plot, from dataset {\tt
    \small  2011.0.00465.S}. Top: cross-correlation
  before outlier filtering.  Middle: after outlier filtering, but
  before astrometric WCS correction. Lower plot: after the alignment
  procedure described in Sec.~\ref{sec:ALMAalign}. In this case, the
  applied WCS offset is (-0.063\arcsec, -0.016\arcsec) along R.A. and
  Dec. We see here that this band~7 dataset {\tt \small \sc
    ALMA\#2011.0.00465.S} is affected by faint outliers, this is why
  we used the data from {\tt \small \sc \small
    ALMA\#2011.0.00318.S} \citep[][]{2013PASJ...65L..14F}. Since the dynamic
  range under the crescent is not limited by thermal noise, we kept
  only the last scheduling block, observed on 10-Aug-2012, to minimize
  pointing offsets. For consistency we tied the astrometric pointing
  to {\tt \small 2011.0.00465.S}.   \label{fig:VVplot}}
\end{figure}

We aligned the multi-frequency interferometer data by
cross-correlating the visibilities, as illustrated in
Fig.~\ref{fig:VVplot}.  We first approximately bring two independent
visibility datasets to a common $uv$-coverage in the following way. We
select as reference the dataset with the more compact $uv$-coverage
(usually the lower frequency), and produce a deconvolved model using
{\tt uvmem} of the visibility dataset with the more extended
$uv-$coverage.  We then compute synthetic interferometer data
corresponding to the more compact $uv-$coverage using our tool {\tt
  uvsim}. For consistency the procedure is also applied to the
reference. Having produced comparable visibility datasets, we
regularly sample shifts in the world coordinate systems (WCS) of the
two frequencies in a uniform grid of positional offsets
$\{\vec{\delta}_i\}$ (assuming that the parallactic calibration is
perfect in both cases).For each WCS offset $\vec{\delta}_i$ we
computed shifted band~9 visibilities
\[
V^\mathrm{shift}_{b9}(\vec{u}_k)  =  V_{b9}(\vec{u}_k)  \exp\left[i2\pi (\vec{\delta}_i \cdot \vec{u}_k ) \right].
\]
We then calculated a linear regression between real and imaginary
parts of visibilities, used as independent data, and computed the
L$_2$ distance (squared norm) between model and data. We form an image
$\chi^2(\{\vec{\delta}_i\})$ of the L$_2$ distances for each
$\vec{\delta}_i$. The best WCS shift is given by the centroid
$\vec{\delta}_\circ$ of an elliptical fit to
$\chi^2(\{\vec{\delta}_i\})$. Although we measure the shifts
$\vec{\delta}_\circ$ in the $uv-$plane, the actual application of
$\vec{\delta}_\circ$ to bring images to a common WCS system is carried
out after image restoration, so in the image plane.

\subsection{Spectral index maps} \label{sec:alpha}

\begin{figure*}
  \begin{center}
    \includegraphics[width=0.26\textwidth,height=!]{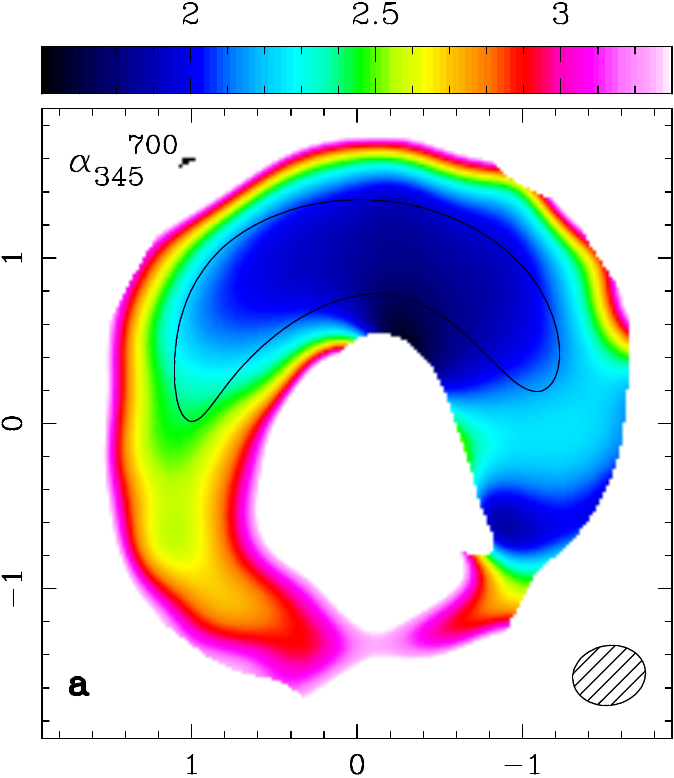}
    \includegraphics[width=0.72\textwidth,height=!]{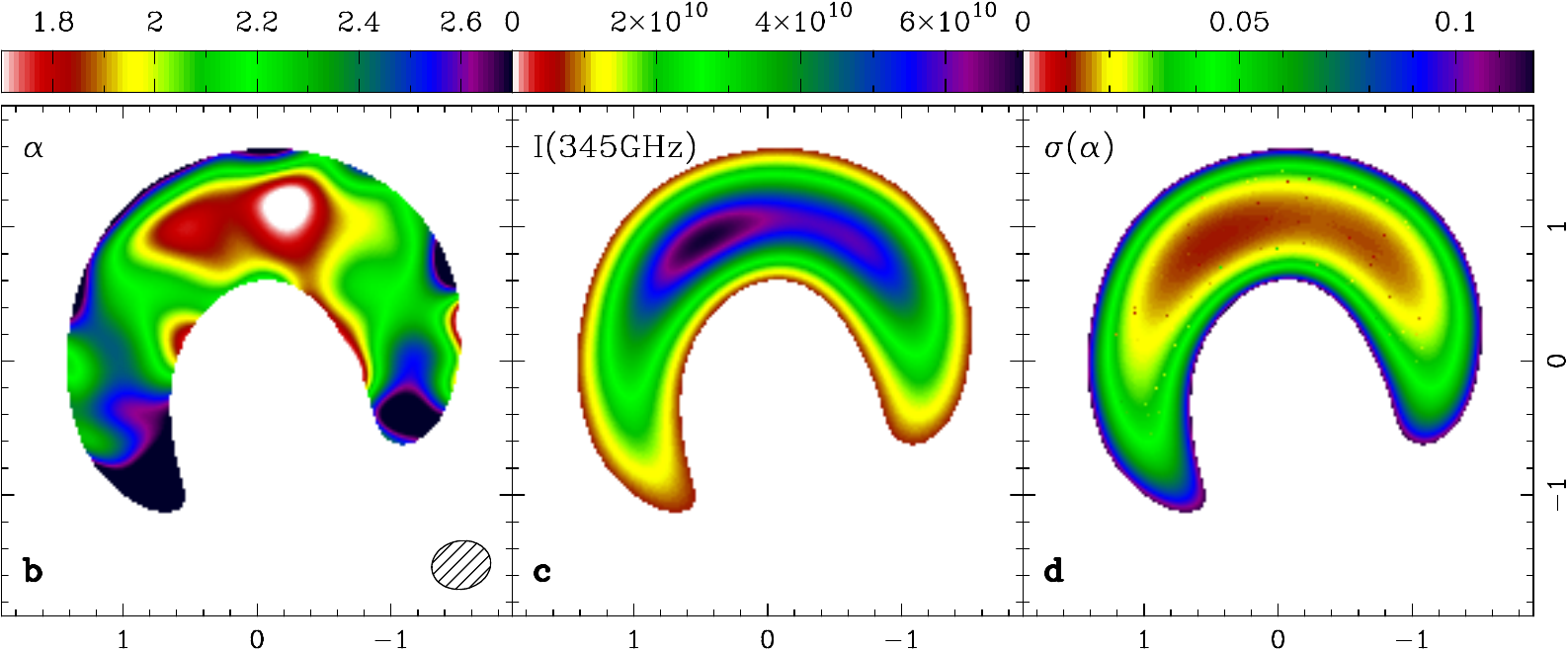}
    \end{center}
\caption{\small Spectral index map from the aligned images of the ALMA
  Band~7 and Band~9 datasets.  {\bf a)} spectral index from the
  comparison between band~7 and band~9, $\alpha_{345}^{700}$.  The
  single contour shows the Band~7 continuum traced at
  half-maximum. {\bf b-d)} Power-law fits to the intra-band~7 ALMA
  data, using dataset {\tt \small \sc \small
    ALMA\#2011.0.00318.S} \citep[][]{2013PASJ...65L..14F}. {\bf b)} spectral
  index map $\alpha(\vec{x})$. {\bf c)} best-fit amplitude at the
  reference frequency of 345~GHz, in Jy/sr. {\bf d)} thermal
  uncertainty on the spectral index map. \label{fig:alphab7b9}}.
\end{figure*}

Thanks to the positional alignment procedure, we can build a
Band~7/Band~9 spectral index map by bringing both datasets to an
approximately common $uv$-coverage, as outlined in
Sec.~\ref{sec:ALMAalign}. We then compute restored images of the {\tt
  uvmem} models.  The resulting Band~7/Band~9 spectral index map is
shown in Fig.~\ref{fig:alphab7b9}. We caution that the absolute values
of the spectral index map $\alpha$ is dependent on the chosen absolute
flux calibration and SED, as explained in
Sec.~\ref{sec:absoluteSED}. However, the relative variations in
$\alpha$ are significant, and independent of the chosen
scale. Fig.~\ref{fig:alphab7b9} corresponds to the flux scale labelled
{\em C13scale} (see Sec.~\ref{sec:absoluteSED}).

% \textcolor{red}{*** perhaps connect
%  Fig. 2, on the fixed-beta T,tau maps from the intra-band7/band9
%  fits, with this spec index map?  ***}
%
%, and 0.05\arcsec~between Band~7 and Band~6

%~/common/genfig_muppet/specindex_b9_b7.pl
%~/common/T_map/fit/specindex_map_nolines_F14/genfig_plaw.pl

In order to build a spectral index map based on the intra-band
frequency lever of the ALMA band~7 data, we selected the lowest
frequency spectral window as reference $uv$-coverage. For instance in
the spectral fits to the intra-band spectral windows (SPWs) of the
ALMA band~7 data {\tt \small \sc \small ALMA\#2011.0.00318.S}, this
corresponds to 329.3~GHz. We then bring all 4 spectral windows in
band~7 to a comparable $uv$-coverage, by following the procedure
outlined in Sec.~\ref{sec:ALMAalign} (so the reference SPW was also
processed in the same way as the other 3 SPWs).

Once all four spectral windows are brought to a comparable
$uv-$coverage, we extract a spectral index map by fitting the restored
specific intensity maps with a power-law. There are therefore two free
parameters per pixel, the intensity at the reference frequency, and
the power-law exponent. Fig.~\ref{fig:alphab7b9} shows the resulting
fit, along with the 1~$\sigma$ error on the spectral index. We caution
that the absolute values of this intra-band spectral index map depends
on the choice of SED scale, as explained in
Sec.~\ref{sec:absoluteSED}. However, the morphological trends are
independent of the total flux density scale.

\subsection{Alignment of the ATCA and ALMA data} \label{sec:ATCAALMAalign}
The offset between the ATCA and the ALMA Band~7 data from {\tt \small
  \sc \small ALMA\#2011.0.00465.S} was within our measurement errors.
Since in the ATCA data we have two IFs centered on 33~GHz and 35~GHz,
we can test the alignment by cross-correlating each IF separately, as
in Sec.~\ref{sec:ALMAalign}, and also by requiring consistency between
real and imaginary parts. It turned out that the ATCA/Band~7 offset is
systematically smaller than 0.15\arcsec, and changes sign. We
concluded that the ATCA and ALMA pointing are consistent, but that the
morphologies are too different for a more refined alignment with our
cross-correlation scheme.
We also attempted the standard strategy of identifying a common anchor
point, in this case the stellar position. We document this exercise as
it bears on the morphology of the signal from the star's vicinity in
ALMA band~7. The central free-free signal, at the limit of dynamic
range in band~7, is difficult to use as astrometric reference as its
morphology is sensitive to self-calibration, and varies from a compact
inner disk reported by \citet[][]{2013PASJ...65L..14F} to a
gap-crossing filament in our previous
analysis \citep[][]{Casassus2013Natur}. However, the continuum filament we
image in our self-calibrated band~7 data, after filtering for outliers
using our tool {\tt uvreweight}, culminates in a local maximum near
the central star, now in closer agreement with  \citet[][]{2013PASJ...65L..14F}.  Given the uncertainties affecting the
very faint stellar signal in band~7, the default astrometries for ATCA
and ALMA band~7 {\tt \small \sc \small ALMA\#2011.0.00465.S} appear to
be consistent.
We note, however, that the 11.5h emission appears to extend outwards
in radius, by about $0.3\pm0.1$\arcsec~ compared to ALMA band~7, while
the 1h clump does not. But given the above pointing uncertainties,
this effect is not significant enough to warrant further consideration
and modeling.
We brought the multi-frequency data of HD~142527 to a common
$uv-$coverage following the same procedure as for the multi-frequency
ALMA data. The main difference is that here we used the ATCA data as
reference. The resulting restored images are shown in
Fig.~\ref{fig:ATCAref_RGB}a, where the origin of coordinates
corresponds to the location of HD~142527 at the Jun 2012 epoch. The
central point source at 34~GHz is satisfactorily aligned with the
origin.

%
%\begin{figure}
%\centering
%\includegraphics[width=0.6\textwidth,height=!]{contmultif_uvmem_withaxes.pdf}
%\caption{RGB image summarizing the alignment of the multi-frequency
%  continuum data, so after filtering for the ATCA response. In red we
%  show the ATCA image, band~7 is in green, and band~9 is in blue. The
%  contours follow the ATCA image, at fractions of 0.4 and 0.7 of the
%  peak intensity.   \label{fig:ATCAref_multif}}.
%\end{figure}

\section{Monte-Carlo tests for the multi-frequency morphological differences} \label{sec:MCtests}

Images obtained with pre-ALMA radio-interferometry are known to be
affected by $uv$-coverage artifacts, especially with ATCA. Despite the
efforts invested in covering the $uv$-plane in several ATCA runs, with
different array configurations, it is important to test for such
$uv$-coverage artifacts, that could for instance inject the 11.5h
local maximum.  We first produce deconvolved model images of the ALMA
band~7 and band~9 visibility data using our tool {\tt uvmem}, and then
simulate ATCA observations on these model images, following exactly
the same $uv$-coverage as in the ATCA 34GHz concatenated data using
our tool {\tt uvsim}. The simulated visibilities are then scaled by a
linear regression (forced to cross the origin), so that the simulated
and ATCA visibilities are comparable in magnitude. Following this
scaling, we inject thermal noise given by the observed visibilities
weights (having first applied our tool {\tt uvreweight} to replace the
ATCA visibility weights by the observed scatter). We repeat this
procedure 200 times before taking statistics.

\begin{figure*}
  \begin{center}
    \includegraphics[width=0.9\textwidth,height=!]{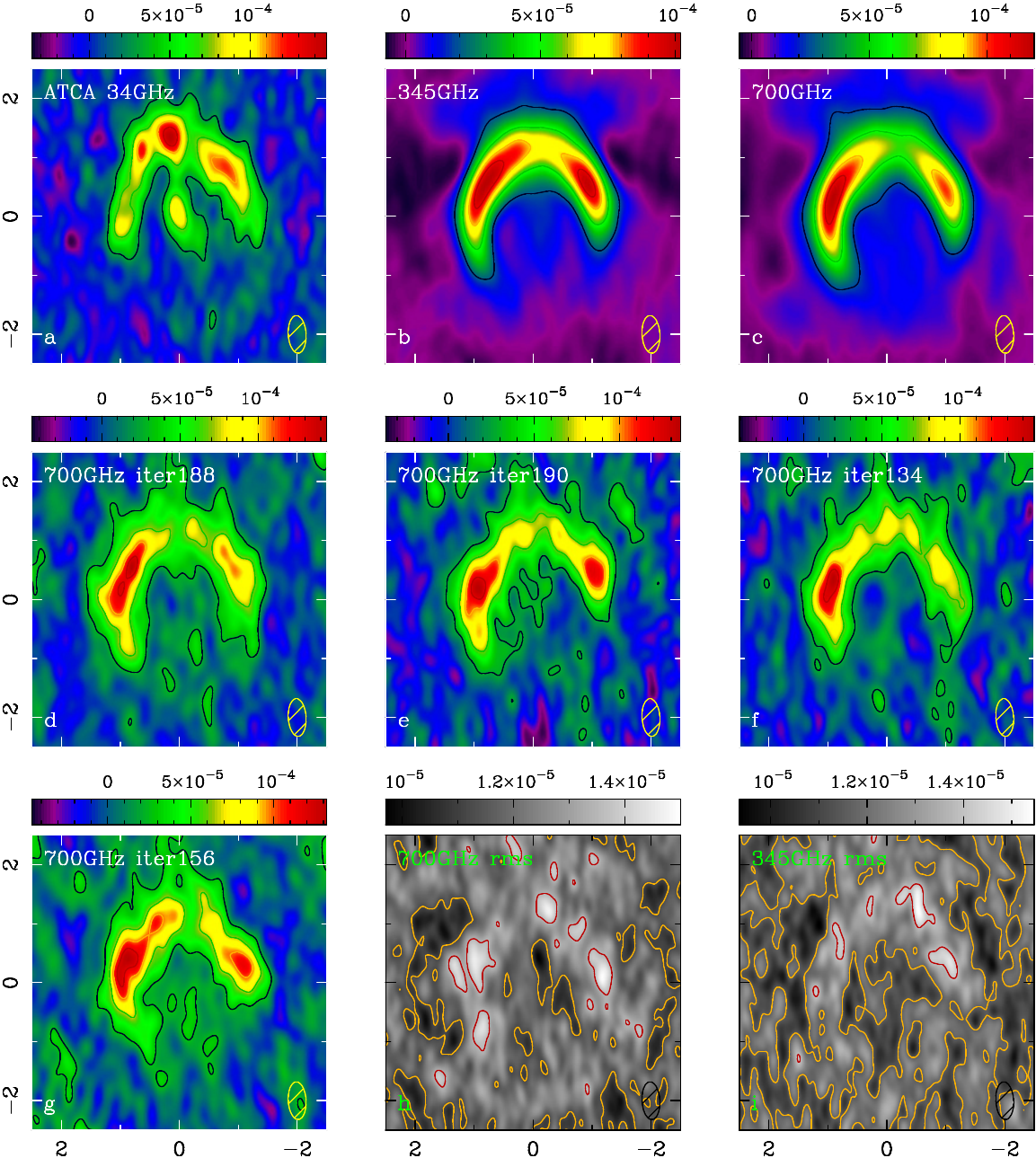}
    \end{center}
  \caption{\small Statistics of the multi-frequency morphological
    differences inferred from Monte-Carlo (MC) simulations. All images
    have been restored from {\tt uvmem} models.  \textcolor{blue}{\bf
      a)}: restored ATCA image at 34~GHz.  \textcolor{blue}{\bf b)}:
    average of ATCA observations with 200 realizations of noise on the
    band~7 data. \textcolor{blue}{\bf c)}: average of the MC
    simulations for band~9 \textcolor{blue}{\bf d)-g)}: different
    realizations of noise.  \textcolor{blue}{\bf h)}: rms scatter of
    the band~9 simulations. \textcolor{blue}{\bf i)}: rms scatter of
    the band~7 simulations.    \label{fig:MCtests}}
\end{figure*}

Fig.~\ref{fig:MCtests} summarizes the result of these Monte-Carlo
simulations.  We see that the signal is more extended in the ALMA
bands than it is in ATCA, so that the visibility scaling we applied
results in lower signal-to-noise ratio than in the actual ATCA
observations. In this sense our MC simulations are conservative, we
are injecting more noise relative to the signal (by about 40\% in
band~7). The final rms dispersions are up to $\sim$1/7 of the
differences between the ATCA and ALMA images - in other words, the
differences between ATCA the ALMA images range from $-$3$\sigma$ to
+6.2$\sigma$. The reduced $\chi^2$ taken over pixels where both ATCA
and ALMA are above 1/3 of the peak, is 4.2 for band~7, and 4.9 for
band~9, for 12 independent data points (the number of ATCA beams
enclosed in the 0.3 times maximum contour) - the images are thus
different with a very high confidence level. In particular, under the
11.5h clump (so at the ATCA maximum), the difference with the band~7
average is 5.9 times the band~7 rms.

%~/common/genfigs_muppet/summary_multif_cont.pl

An additional test was performed by subtracting the synthetic band~7
ATCA visibilities from the 34~GHz ATCA data, after appropriate scaling
by linear regression. The stellar emission stands out in the
residuals, as well as the 11h ATCA clump. These significant residuals,
both positive and negative, point to genuine morphological differences
between 34~GHz and 345~GHz. In particular, the peak residual
corresponding to the 11h clump is at 50~$\mu$Jy~beam$^{-1}$, while the
thermal noise is at 10~$\mu$Jy~beam$^{-1}$.

\section{Coincidence between ATCA maxima and  molecular line decrements} \label{sec:decrements}

\begin{figure}
  \begin{center}
    \includegraphics[width=0.75\columnwidth,height=!]{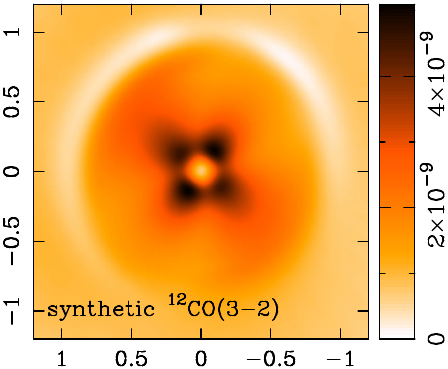}
    \end{center}
\caption{\small Frequency-integrated image of the $^{12}$CO(3-2)
  emission calculated with RADMC3D on our preferred model for the dust
  trap of HD~142527. $x-$ and $y$- axis indicate offset from the
  stellar position along R.A. and Dec., in
  arcsec. \label{fig:decrement_synthetic}}
\end{figure}

The 34~GHz clumps coincide with the molecular-line decrements in
HCO$^+(4-3)$ and in CO$(3-2)$ that have been previously
documented \citep[][]{Casassus2013Natur}. It has been suggested that an
optically thick continuum may somehow be responsible for the
correlation of the $^{12}$CO decrements with the 345~GHz
crescent \citep[][]{Casassus2013Natur,vdP_2014ApJ...792L..25V}. Indeed, as
shown in Fig.~\ref{fig:decrement_synthetic}, the synthetic dust trap model
documented below readily reproduce such decrements with a dust to gas
mass ratio of 94, a standard CO abundance of $10^{-4}$, and the
following temperature profile:
\begin{eqnarray}
  T & = &  70~\rm{K}, ~\rm{if}~ r < 30~\rm{AU}, ~\rm{or} \\ \nonumber
  T & = & 70~\rm{K} (r/ (30~\rm{AU} )^{-0.5} ~\rm{if} ~ r \ge 30~\rm{AU}.  
\end{eqnarray}
However, setting the gas temperature to a representative dust species
in the outer disk (we chose the average population of amorphous carbon
grains) attenuates the depth of the decrements - they are restricted
to the outline of the underlying optical depth field. The decrements
are reproduced by the above prescription for the gas temperature
because the gas temperatures in the crescent, at $r=140~$au, are
$\sim$30~K, and cooler than the continuum temperatures in the model,
which reach 50 to 60~K. Thus in this case the gas acts like a cold
foreground on a hotter continuum background, which after continuum
subtraction appears like foreground absorption.  In fact, CO at higher
altitudes could be hotter than the background continuum, so could
still stand out after the subtraction of a colder optically thick
continuum. Yet the gas temperature prescription above is inspired by
the molecular line data, and higher temperatures can be ruled out as
they greatly overestimate the integrated CO line fluxes.

%
%
%\newpage
%\begin{figure}
%  \begin{center}
%    \includegraphics[width=0.8\textwidth,height=!]{fig_decrements.pdf}
%    \end{center}
%\caption{\small Molecular decrements under the 34~GHz emission clumps
%  {\bf a} 34~GHz restored image overlaid on the HCO$+$ peak intensity
%  map from MEM models. Note the coincidence of decrements in HCO$^+$
%  with the 34~GHz peaks. {\bf b} 345~GHz continuum in blue contour
%  compared with the CO(3-2) peak intensity map (MEM, velocities with
%  line-of-sight diffuse CO have been excluded). \label{fig:decrements}}
%\end{figure}
%

%\textcolor{red}{The flux density in the F14 data delivery is 2.0 at
%  336~GHz, but is reported to be 2.7 in F14, while I get 2.45 from my
%  selfcal of their visibility data, and the value in C13 is 3.52~Jy at
%  350~GHz}.

%, but should not impact the morphology of relative
%variations in $\alpha$, and the trend for a minimum under the
%continuum crescent
%
%Rather than to quantify the observational errors,
%
%The
%existence of solutions for the grey-body fits (3 parameters for 3 data
%points) only loosely limited the range of possible absolute flux
%values.

%The recovered flux density for HD~142527 is 8~Jy at 702.4~GHz which
%may climb to 10~Jy including some amount flux loss, which we estimate
%at about 20\% by simulating observations on synthetic band~9 images.
%By comparison, t

%These uncertainties affect the absolute values of the
%spectral index maps (although not the morphology of the optical
%depth).

\section{Absolute flux scale} \label{sec:absoluteSED}

The inferred grey-body diagnostics and spectral index maps depend on
the absolute calibration of the input multi-wavelength data. There are
however substantial uncertainties on the observed SED. For instance,
the flux densities reported for different ALMA programs in band~7 vary
by up to 50\%,  \citep[][]{Casassus2013Natur,2013PASJ...65L..14F}.  The
band~9 flux densities are affected by similarly large
uncertainties. The {\em Herschel} SPIRE flux
density \citep[][]{vanderwiel2014MNRAS.444.3911V} at the same frequency is
13.7~Jy. We investigated two plausible scenarios for the observed
SED. The pectral trends reported here are derived from the first of
these flux scales, which we label {\em C13scale}, where we adopted a
reference band~7 flux density of 3.5~Jy at 345~GHz, and took the SPIRE
flux density for band~9. The comparison between the SPIRE flux density
and the band~7 levels implies a spectral index $\alpha_{b7}^{b9} \sim
1.9$. On the other hand, the band~6 ALMA data at
250~GHz \citep[][]{Perez2015ApJ...798...85P} yields $\alpha_{b6}^{b7} \sim
2.7$. We thus assumed that a plausible value for the intra-band~7
spectral index is $\alpha_{b7}=2.2$.

The synthetic dust trap, documented below, fits the SED within the
error bars. So, for the purpose of comparing the observations with the
synthetic dust trap predictions, we also considered a second flux
sale, which we labelled {\em SEDscale}. This scale has a rather high
band~9 flux density of 15.7~Jy at 700~GHz, and is rather low in
band~7, with 2.17~Jy at 342.9~GHz, while it comes very close to the
reported flux density at 34~GHz, with 1.34~mJy. All morphological
trends are similar in both flux scales, but the highest optical depths
in the {\em SEDscale} are lower - dropping by about 2.5 in {\em
  C13scale} to 0.7 in {\em SEDscale} (at 345~GHz). The temperature map
appears to be more irregular in {\em SEDscale}, which suggests that
{\em C13scale} may be more realistic.

\section{Line of sight temperatures and optical depths} \label{sec:Ttaubet}

\begin{figure*}
\begin{center}
\includegraphics[width=\textwidth,height=!]{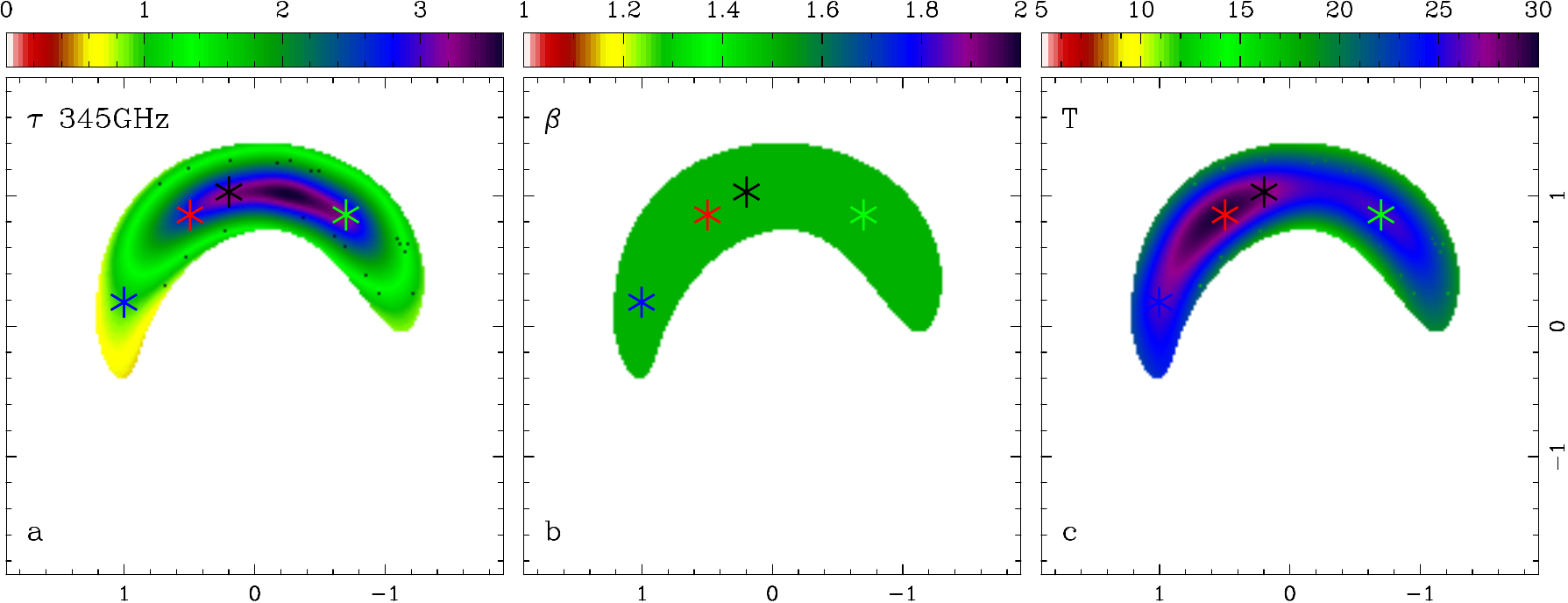}
\includegraphics[width=0.8\textwidth,height=!]{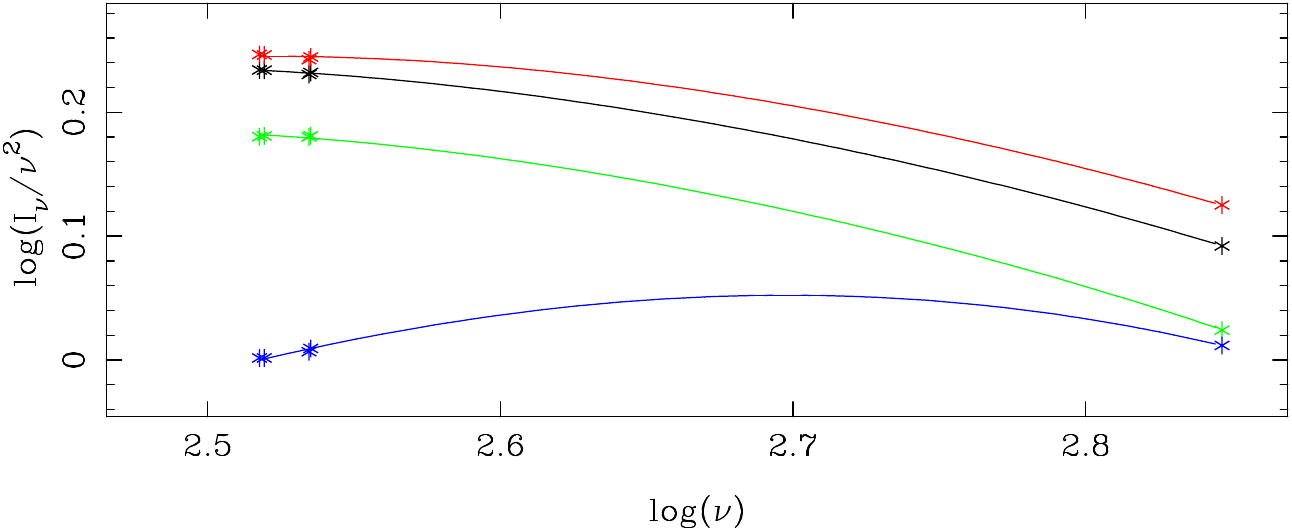}
\end{center}
\caption{ \small Spectral properties of the sub-mm continuum, obtained
  by comparing the intra-band ALMA band~7 spectral information with
  ALMA band~9, and with flux-density scale labelled {\em C13scale}.
  {\bf Upper panel}: {\bf a} optical depth map at 345~GHz,
  $\tau_\circ(\vec{x})$ {\bf b}: flat emissivity law, taken here as
  $\beta = 1.5$ - this plot allows to better locate the asterisks that
  indicate the locations of the example spectra {\bf c}: temperature
  map $T_s(\vec{x})$.  {\bf Lower panel:} example SEDs extracted at
  selected positions, indicated by asterisks in the upper panel. The
  $y$-axis shows specific intensity normalized by a $\nu^2$ spectrum,
  in arbitrary units. \label{fig:Ttaubet_b9}}.
\end{figure*}

\begin{figure*}
\begin{center}
\includegraphics[width=\textwidth,height=!]{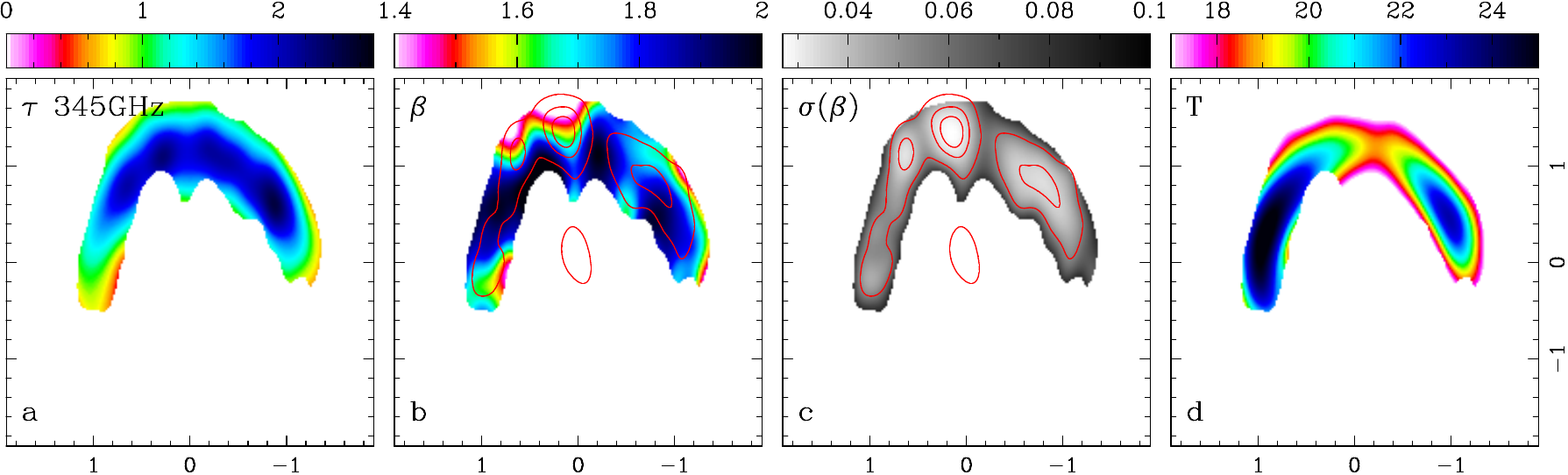}
\includegraphics[width=\textwidth,height=!]{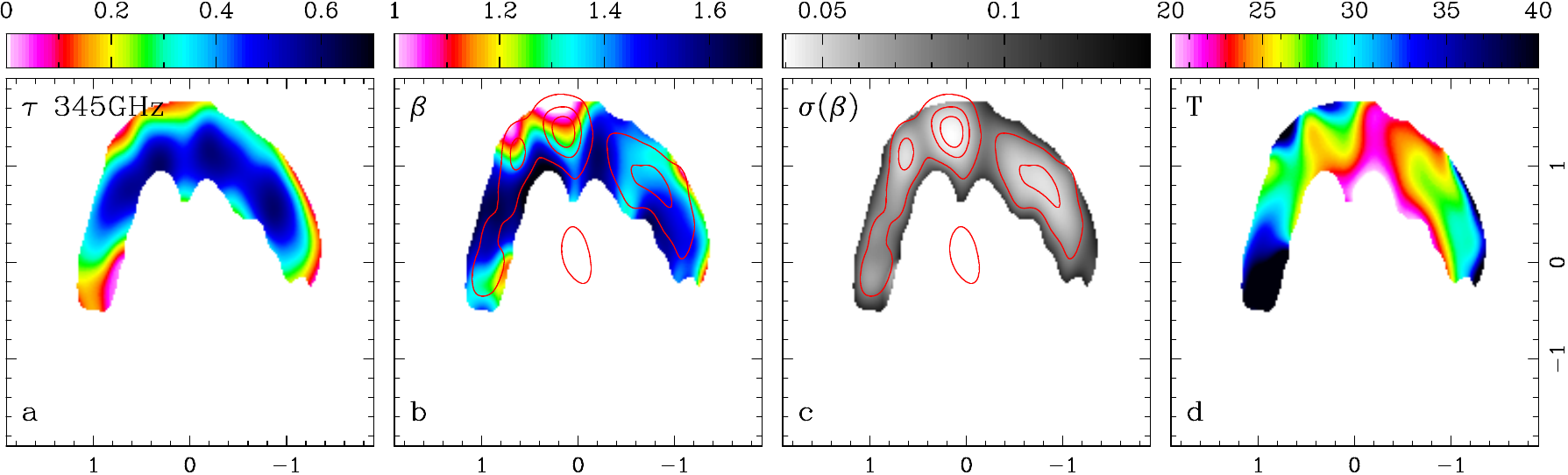}
\end{center}
\caption{\small Optical depth maps derived from the ATCA/ALMA
  multi-frequency data. The emissivity minimum at 11.5h indicates the
  presence of larger grains. The \textcolor{blue}{\bf upper panel}
  corresponds to the absolute flux density scale labelled {\em
    C13scale} in Sec.~\ref{sec:absoluteSED}, with a high flux density
  in band~7 (see Sec.~\ref{sec:absoluteSED}), while the
  \textcolor{blue}{\bf lower panel} corresponds to the flux scale in
  our preferred RADMC3D model, labelled {\em SEDscale} (see
  Sec.~\ref{sec:absoluteSED}).  {\bf a}: optical depth map at the
  reference frequency of 345~GHz. {\bf b}: line of sight emissivity
  index map $\beta_s(\vec{x})$, with ATCA specific intensity contours
  in red.  {\bf c}: root-mean-square uncertainties on the emissivity
  index map.  {\bf d}: line of sight temperature,
  $T_s(\vec{x})$. \label{fig:Ttaubet} }
\end{figure*}

The grey-body uniform-slab models are used as diagnostics for the
optical depth, and are meant to characterize the emergent spectrum at
a given line of sight with three parameters, adjusted to three
frequency points.  By applying the grey-body diagnostic to the ALMA
band~7 amplitude and slope, and including band~9, we find that the
crescent is optically thick in band~7.  Fig.~\ref{fig:Ttaubet_b9} shows
that grey-body line-of-sight models provide a fit to the observations,
despite the use of a constant emissivity law $\beta_S(\vec{x}) =
1.5$. We used the {\em C13scale} flux-density scale; {\em SEDscale}
produces similar maps but the low band~7 flux density results in lower
optical depths. We also applied the grey-body diagnostic to the
ATCA/ALMA multifrequency data, as summarised in Fig.~\ref{fig:Ttaubet}. The
temperature map is essentially fixed by the band~9 spectral point. We
note a minimum in temperature towards 0.5~h, which is also seen when
using the ALMA data alone. The similarities between both temperature
maps in {\em C13scale} is surprisingly good, considering the
differences in $uv$-coverage, and the different treatments for
$\beta$.

%bol04:41:22~/common/T_map/fit/scale_SED_wb9_nolines_F14.scaleRADMC.fixbeta$ rsync -va exampleseds.pdf  ~/common/muppet/exampleseds_F14_wb9_fixbeta.pdf

%bol01:23:01~/common/T_map/fit/ATCAfilt_uvmem$ rsync -va fig_Ttaubet_errors.pdf  ~/common/muppet/fig_Ttaubet_ATCAfilt_uvmem_errors.pdf
%bol09:47:27~/common/T_map/fit/ATCAfilt_uvmem_RADMC3Dscale$ rsync -va ./fig_Ttaubet_errors.pdf  ~/common/muppet/fig_Ttaubet_ATCAfilt_uvmem_errors_radmcscale.pdf

The uncertainties on the opacity index $\beta_S(\vec{x})$ were
estimated by varying the ATCA intensities by $\pm 1\sigma$, where
$\sigma$ is the scatter in the dirty map of image synthesis
residuals. The resulting uncertainty map is shown in
Fig.~\ref{fig:Ttaubet}. The peak at 34~GHz, under the 11.5h clump,
corresponds to $\beta_S = 1.18 \pm 0.04$ (in the {\em SEDscale},
similar significance values are obtained in the {\em C13scale}),
whereas the average value outside the regions of low $\beta_S$ is
1.50, with a rms dispersion of 0.07 and typical values varying from
$1.41\pm0.05$ at 1h, up to $1.63\pm0.08$ at 9.5h. Compared to the
average value of 1.50, the minimum in $\beta_s$ under the 11.5h clump
is significant at $\sim$7$\sigma$.

%The error bars on the multi-frequency data
%$I_{\nu}(\vec{x})$ were obtained by measuring the root mean squared
%dispersion of the dirty-maps in the image-synthesis residuals (with
%the {\tt uvmem} image model).
%

%For DEV shifts ATCA:
%DEV
%gsig:
%A ( 60.1, 103.1) = 3.91e-02 
%A: 15:56:41.8906  -42:19:22.327  DEG: 239.174544066073  -42.3228685040339 
%beta:
%A ( 55.0, 112.6) = 1.21e+00 
%A: 15:56:41.8908  -42:19:22.237  DEG: 239.174544921267  -42.322843531443 
%
%
%gsig:
%A ( 107.3, 76.4) = 4.66e-02 
%A: 15:56:41.8055  -42:19:22.860  DEG: 239.174189790287  -42.3230166700518 
%beta:
%A ( 102.3, 81.2) = 1.44e+00 
%A: 15:56:41.8054  -42:19:22.864  DEG: 239.174189353293  -42.3230178985932 
%
%
%

%Since here the data have been filtered to the ATCA $uv-$coverage, the
%differences between both temperature maps probably reflect some level
%of flux-loss: missing Fourier components in the ATCA-filtered maps
%lead to a somewhat deeper temperature decrement due North, where the
%source is extended in the direction of the minor axis of the synthetic
%beam.
%

\section{Synthetic dust trap model} \label{sec:dtrap}

We constructed a synthetic model of a protoplanetary disk with a dust
trap to make predictions comparable to the ALMA and ATCA observations
of HD~142527.  This synthetic model matches the observed SED (see
below), and is a perfected version of that which accounts for the
scattered-light shadows \citep[][]{Marino2015ApJ...798L..44M}, with a
tilted inner disk.  This model allows us to 1- assess if the
observations can be interpreted with steady-state dust trapping
formulae, given the tilted-inner disk, and 2- estimate the biases
involved in the uniform-slab grey-body model.

The outer disk is where the dust trap is observed, so we describe it
in more detail.  It goes from 115 au to 300 au with a rounded disk
wall between 115 to 140 au, and is composed of 5.0$\times10^{-5}$
M$_{\odot}$ in amorphous carbon grains and of 2.6 $\times10^{-3}$
M$_{\odot}$ in silicate grains mixed with ices (using the
Maxwell-Garnett rule). To simulate the dust trap effect on different
grain sizes we used the radial trap model from
\citet[][]{Pinilla2012A&A...538A.114P} coupled with the prescriptions
for azimuthal trapping from
\citet[][]{Birnstiel2013A&A...550L...8B}. We first define an
axisimmetric gas distribution with a rounded disk wall, as previously
described in \citet[][]{Marino2015ApJ...798L..44M}, but with adjusted
parameters $\gamma=7$, $w=0.13$ and $r_c=155.0$ au. We compute a
critical grain size \citep[][]{Pinilla2012A&A...538A.114P} $a_c$ at
133 au for which bigger particles would be radially trapped in the
pressure maximum, while smaller particles would follow the gas density
background. Using a total gas mass of 0.25 $M_{\odot}$ and an
$\alpha_\mathrm{t}$ viscous parameter of $10^{-2}$ we obtain a
critical grain size of 0.7~cm at 133~AU. The Stokes number for 0.7~cm
grains, at peak gas densities of 2.6~10$^{10}$~cm$^{-3}$ and 20~K, is
$\sim$0.02 and close to $\alpha_t$.

%described by the following surface density:
%\begin{eqnarray}
%\bar{\Sigma}_{g}(r<r_c)&=&\Sigma_{c} \left(\frac{r}{r_{c}} \right)^{-\gamma}\exp\left[ - \left(\frac{1-r/r_c}{w} \right)^3 \right], \\
%\bar{\Sigma}_{g}(r\ge r_c)&=&\Sigma_{c} \left(\frac{r}{r_{c}} \right)^{-\gamma},
%\end{eqnarray}

%, using Eq.~11 in  \citep[][]{Pinilla2012A&A...538A.114P}

%
%\begin{figure}
%\begin{center}
%  \includegraphics[width=0.6\textwidth,height=!]{a_c.pdf}
% \caption{\small Critical grain size for trapping, $a_c$, for the
%   conditions in the synthetic dust trap model that approximates
%   HD~142527. We see that for radii corresponding to the outer disk,
%   $a_c > $1mm, supporting the lack of obvious trapping in the
%   submm. We chose to fix $a_c = 0.7$cm, corresponding to 133AU.
%   \label{fig:a_c}}
%\end{center}
%\end{figure}

%
%\begin{eqnarray}
%\Sigma_{g}(r,\phi)&=&\bar{\Sigma}_{g}(r)\left[1+A(r)\sin\left(\phi+\frac{\pi}{2}\right) \right], \\
%A(r)&=& \frac{c-1}{c+1}\exp\left[-\frac{(r-R_{s})^2}{2(H)^2}\right],
%\end{eqnarray}

%\begin{equation}
%\rho_{g}(r,z,\phi)=\frac{\Sigma(r,\phi)}{\sqrt{2\pi}H}\exp\left[-\frac{z^2}{2H^2} \right].
%\end{equation}

%, using
%\begin{eqnarray}
%\rho_{d,a}(\phi)=C(a)\rho_{g}\exp \left[ \frac{-\mathrm{St}(\phi,a)}{\alpha_{\mathrm{t}}} \right], \\
%\mathrm{St} = \sqrt{\frac{\pi}{8}} \frac{a}{H} \frac{\rho_s}{\rho_{g}},
%\end{eqnarray}
%where $\mathrm{St}$ is the Stokes number, $a$ is the grain radius,
%$C(a)$ a normalization constant, and $\rho_s$ is their internal mass
%density.

The above radial distribution is modulated in azimuth, using
previously described formulae and
symbols \citep[][]{Birnstiel2013A&A...550L...8B} (here we use $R_s=148$~AU,
and an azimuthal contrast in surface density of $c=20.0$). The
scale-height $H$ is given by
\begin{equation}
H(r)=20.0 \left(\frac{r}{130 \ \mathrm{AU}}\right)^{1.17}.
\end{equation}
For the gas volume density we used the standard vertical Gaussian
distribution. The small grains, i.e.  amorphous carbon and silicate
grains smaller than $a_c$, follow the same density field as the
gas. The azimuthal distribution of big grains is calculated according
to their Stokes number, a parameter to account for the gas turbulence
$\alpha_\mathrm{t}$ and the gas
background \citep[][]{Birnstiel2013A&A...550L...8B}. For their radial
distribution we use a Gaussian parameterisation to account for the
spread caused by turbulence:
\begin{equation}
\rho_{d,a}(r,\phi,z=0) = \rho_{d,a}(\phi)\exp\left[ -\frac{(r-R_s)^2}{2\sigma^2}  \right],  
\end{equation}
with $\sigma(a)=H_r\sqrt{\frac{\alpha_t}{St(a)}}$. In order to compute
the density as a function of height $z$, we use the same scale height
as the gas.  We investigated the implementation of grain settling,
following a parameterisation for the dust scale height
 \citep[][]{ArmitagePhilipJ2010}. However, we found that in order to
reproduce the projection effect at $\sim$1h, settling had to be very
inefficient.  Finally the distribution of big grains is normalized to
the global grain size distribution and total dust mass in big grains.

We used RADMC3D \citep[][]{RADMC3D0.39} to transfer the stellar radiation
through the synthetic disk and predict emergent intensity maps, under
the assumption of passive heating. We assumed a standard power-law
distribution in grain sizes, with exponent $-3.5$.
The outer disk is composed of three different dust species
representing a range in sizes described by $a_{\min}$ and $a_{\max}$:
\begin{itemize}
\item 5.0$\times10^{-5}$ M$_{\odot}$ of amorphous carbon grains with radii $1 < a  < 100~\mu$m with optical constants taken from  \citet{Li_Greenberg_1997A&A...323..566L} 
\item 2.5$\times10^{-3}$ M$_{\odot}$ in  icy silicate grains,  with radii of $100\mu\rm{m} < a < a_c $ and optical constants from  \citet{Henning_Mutschke_1997A&A...327..743H} and  \citet{Li_Greenberg1998A&A...331..291L}
\item 1.0$\times10^{-4}$ M$_{\odot}$ in icy silicate grains, with radii of  $a_c< a < 1 $cm and optical constants from  \citet{Henning_Mutschke_1997A&A...327..743H} and  \citet{Li_Greenberg1998A&A...331..291L} 
\end{itemize}

%This choice of grain properties parameterisation we were able to
%account for the observed SED and ALMA and ATCA reconstructed images.
%

%We implemented the parametrised model in RADMC3D using spherical
%coordinates, with regular spacing for the azimuthal angle and log
%spaced bins for the polar and radial coordinates with more resolution
%near 140 au. We used $ 10^{6}$ cells in total, half for the inner disk
%and gap, and the other half for the outer disk. The number of cells
%along the radial, azimuthal and polar directions was 100.
%

%The implementation can be divided in two: the dust opacity and the
%disk model implemented in a python code Setup.py.
  
\begin{figure*}
  \begin{center}
    \includegraphics[width=0.75\textwidth,height=!]{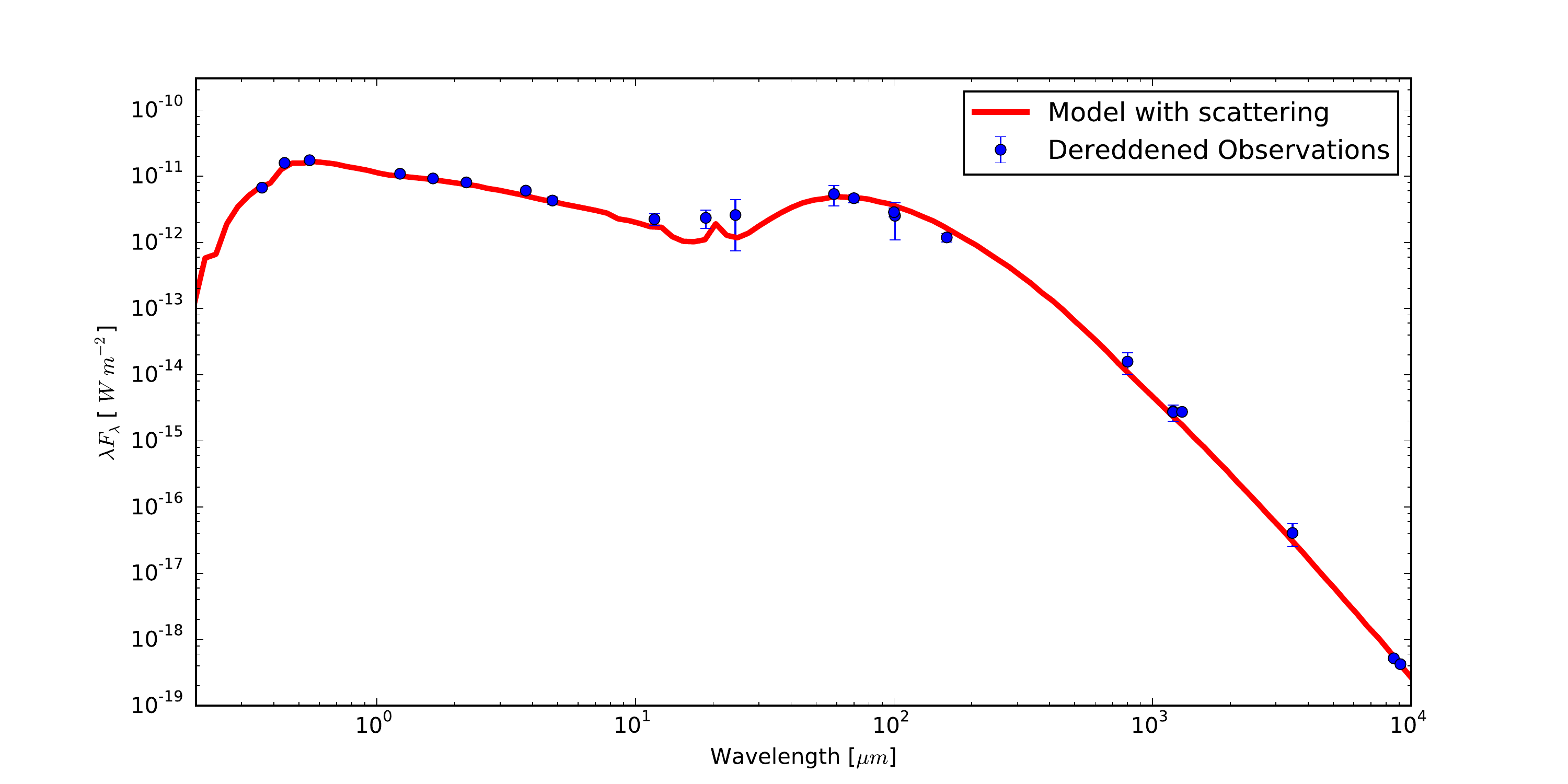}
    \end{center}  
\caption{Observed and model SED for HD~142527. \label{fig:SED}}
\end{figure*}

\begin{figure}
  \begin{center}
    \includegraphics[width=\columnwidth,height=!]{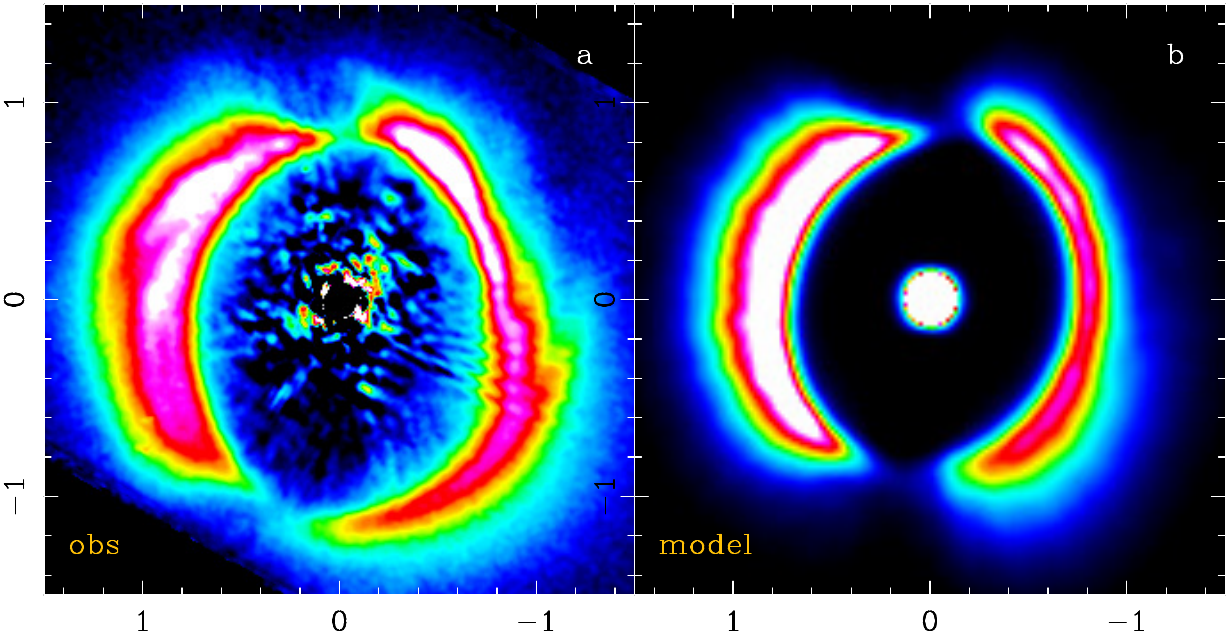}
    \end{center}  
  \caption{Consistency of the synthetic dust trap model with the
    observed scatterd light shadows.  {\bf a:} NACO-PDI H-band image
    from \citet{Avenhaus2014ApJ...781...87A} {\bf b:} Radiative
    transfer prediction in H-band from the dust trap
    model. \label{fig:warpshadows}}
\end{figure}

\begin{figure*}
  \begin{center}
\includegraphics[width=\textwidth,height=!]{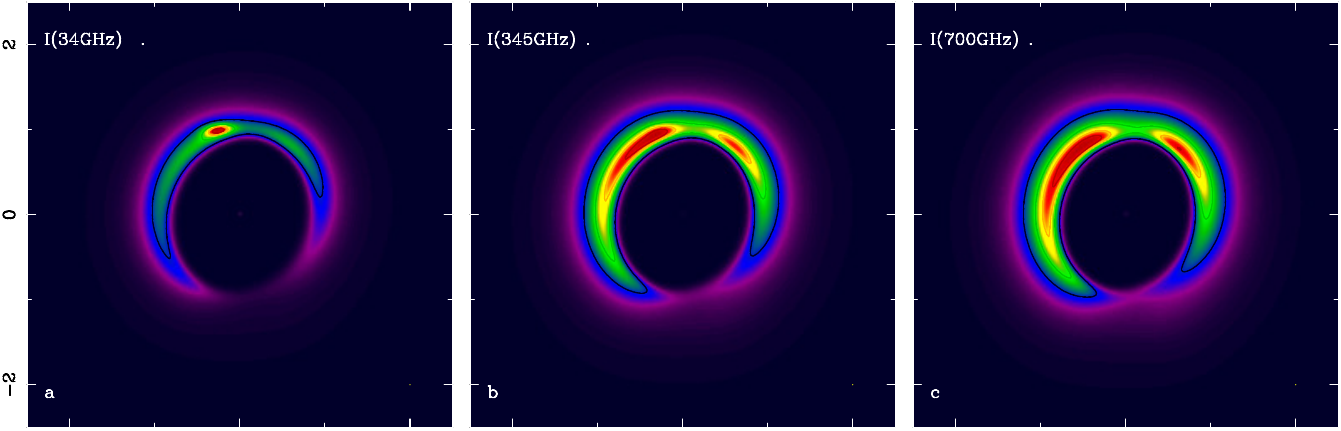} %genfig_images_native.pl
\includegraphics[width=\textwidth,height=!]{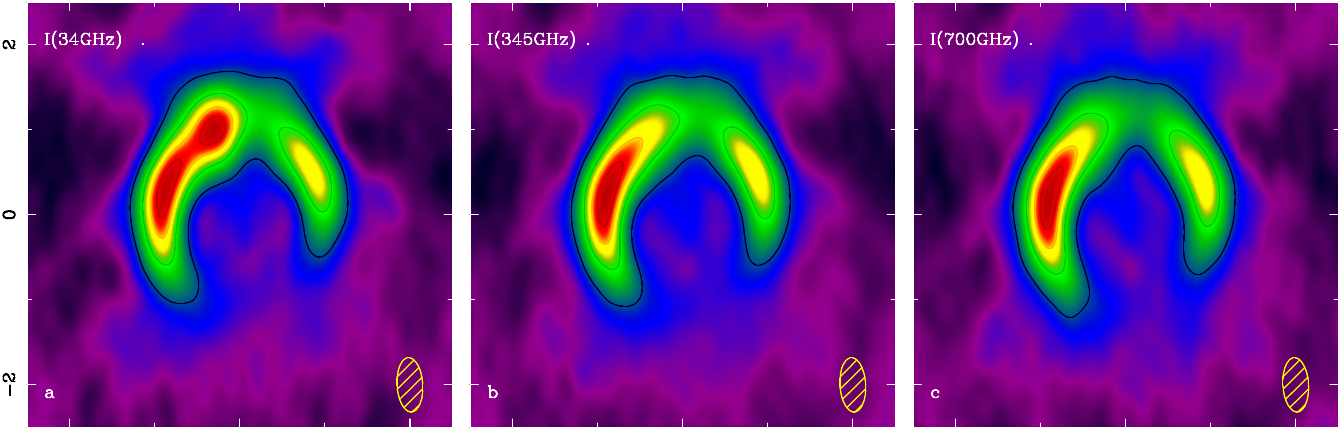} %genfig_images_ATCAfilt.pl
\end{center}
\caption{ Predicted emergent continua at 34~GHz, 345~GHz and
  700~GHz. {\bf Upper panel:} raw RADMC3D prediction. {\bf Bottom
    panel:} predictions filtered by ATCA+{\tt uvmem} response, with
  restoration in natural weights. Note the 11.5h clump of cm-sized
  grains, which stands out at ATCA frequencies, but is absent at ALMA
  frequencies.
\label{fig:radmc_cont}
}
\end{figure*}

As summarised in Fig.~\ref{fig:SED}, the predicted integrated flux
densities fall close to the observed dereddened SED, extracted from
previous SED fitting experiments
\citep[][]{Verhoeff2011A&A...528A..91V,Perez2015ApJ...798...85P}, and
from the {\em Herschel} archive. Fig.~\ref{fig:warpshadows} shows that
the model also reproduces the H-band scattered light images
\citep[this is a perfected version of Fig.~2
  in][]{Marino2015ApJ...798L..44M}.  We switched-off scattering for
the purpose of calculating images at sub-mm frequencies and lower. The
reason for this is that the scattering cross sections calculated using
Mie theory are probably overestimated, given that the actual grains
are probably irregular in shape, or even porous, and so far from the
Mie spheres. The emergent intensity maps in Fig.~\ref{fig:radmc_cont}
show that the 34~GHz emission exhibits a compact clump at the location
of the dust trap, and a more extended crescent that follows the sub-mm
morphology.  In order to simulate the ATCA response, we sampled the
$uv$-coverage of the actual ATCA observations, but on the radiative
transfer predictions, and then averaged 100 {\tt uvmem}
reconstructions with different realizations of noise. This exercise
produced sky maps that are remarkably similar to the ATCA-filtered
observations. We also note that any flux loss is negligible.

%rsyncfromscooby /home/smarino/RADMC/HD142527/Low_ext_wall/sed.pdf  ./data/

%~/common/T_map/dtrap/

\section{Uniform-slab diagnostics on the synthetic dust trap.}
\begin{figure*}
\begin{center}
\includegraphics[width=0.9\textwidth,height=!]{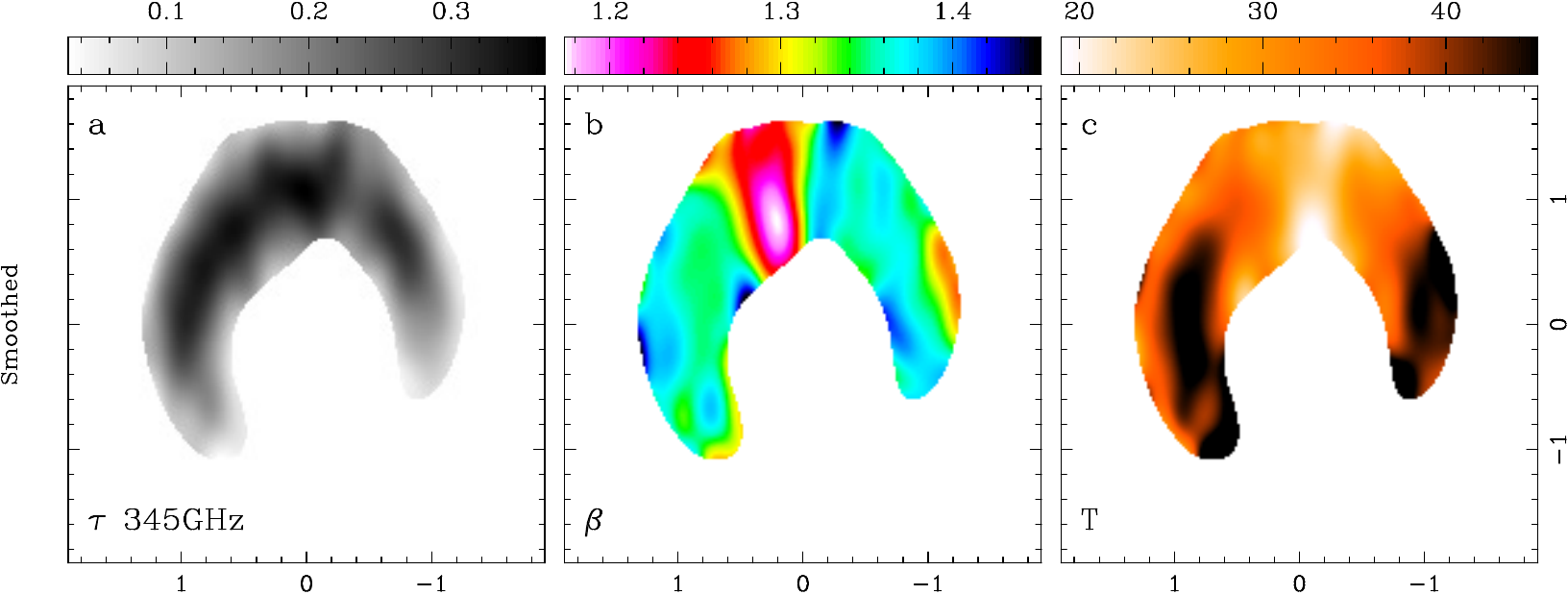} %genfig_Ttaubet.pl
\includegraphics[width=0.9\textwidth,height=!]{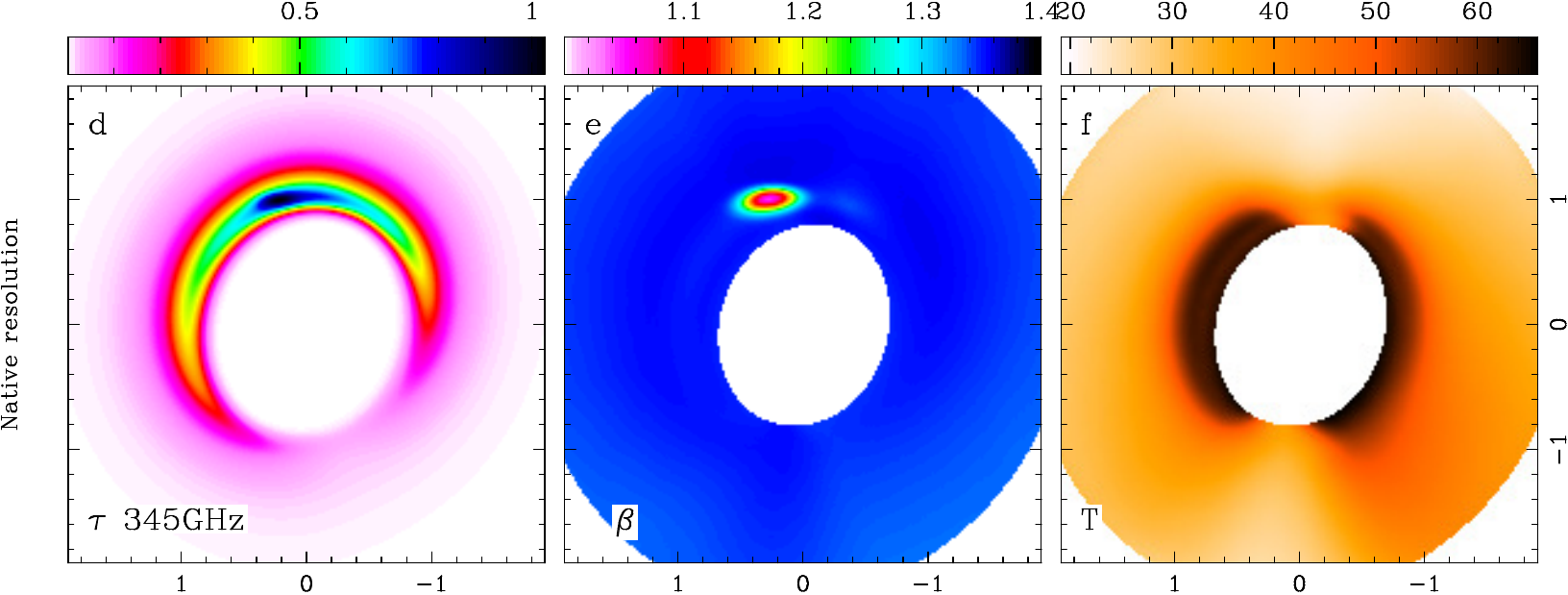} %genfig_Ttaubet.pl
\includegraphics[width=0.55\textwidth,height=!]{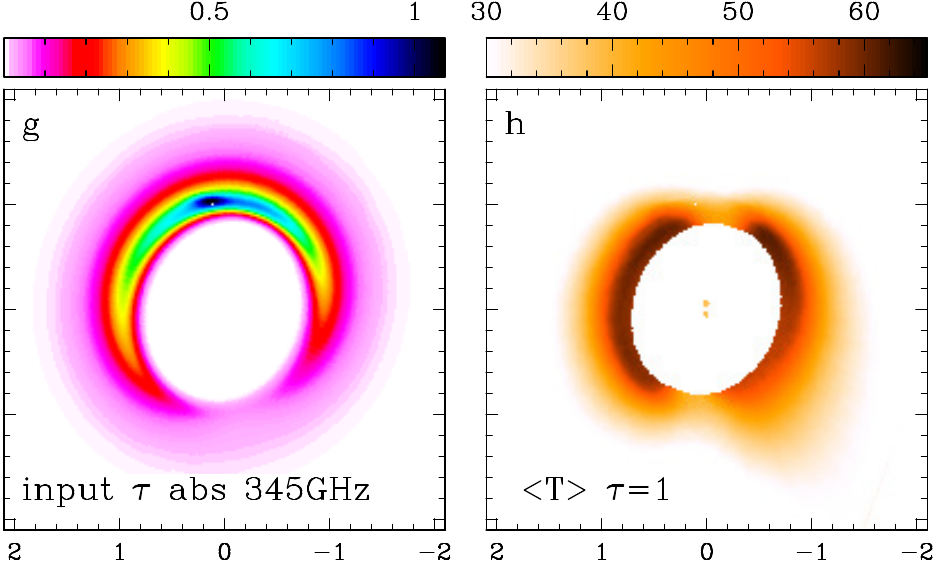}  % ./synthetic_dtrap/rot.pl ./synthetic_dtrap/genfig_maps.pl
\end{center}
\caption{ \small Grey-body 'line-of-sight' diagnostics calculated on the
  radiative transfer predictions, and comparison with input optical
  depth and temperatures. A mask has been applied to compute
  line-of-sight diagnostics at a fraction of 1/5 times peak intensity.
  {\bf a), b), c)}: grey-body diagnostics calculated after filtering
  by the ATCA+{\tt uvmem} response - {\bf a)} shows the optical depth
  map at the reference frequency of 345~GHz, {\bf b)} shows the line of
  sight emissivity index map $\beta_s(\vec{x})$, and {\bf c)} shows
  line of sight temperature, $T_s(\vec{x})$.  {\bf d)}, {\bf e)}, {\bf
    f)}: same quantities, as calculated on the radiative transfer
  predictions. {\bf g)}, {\bf h)}: optical depth and average
  temperature maps used as input for the ray-tracing. See text for the
  definition of average temperature.
\label{fig:Ttaubet_dtrap}
}
\end{figure*}

We also applied the line-of-sight diagnostic based on grey-body models
to the synthetic emergent intensities. As summarised in
Fig.~\ref{fig:Ttaubet_dtrap}, the basic features of the grey-body
diagnostics approximate those seen in the data. There is a mild
extension of the optical depth map into the northern ansa, which could
be emphasized with higher optical depths. The center of the dust trap
coincides with a minimum in $\beta_S$. With this synthetic dust trap
we can also test for biases in the grey-body diagnostics. The input
optical depth map,
\begin{eqnarray}
  \tau(345~\mathrm{GHz}) & =  & \int_0^\infty ds  \rho_g(s) \nonumber \\ 
   & & \int da \sum_i  n_i(a,s) \kappa_i(a,345\mathrm{GHz}) ,
\end{eqnarray}
is remarkably consistent with the grey-body values - at native
resolutions. We can also compare with the grey-body temperature
diagnostic, $T_s({\vec{x}})$, with the line-of-sight average
temperature, as reported by RADMC3D after the thermal solution has
converged,
\begin{eqnarray}
\langle T \rangle  &  =   &   \int_0^\infty ds \rho_g(s) \nonumber \\ 
 & & \int da \sum_i  T(a,s) n_i(a,s) \frac{\kappa_i(a,345\mathrm{GHz})}{\tau(345\mathrm{GHz}}. 
\end{eqnarray}
Fig.~\ref{fig:Ttaubet_dtrap} shows that the two temperature
diagnostics are very nearly equal (to within 10\%). Limiting the
integral to the depth $s$ corresponding to $\tau(345~\mathrm{GHz}) =
1$ resulted in essentially identical maps.

%Finally, we also repeated the exercise of fitting a grey-body to
%predictions for the intra-band~7 `spectral windows' and band~9.  The
%result is shown in \ref{fig:Ttaubet_f14_b9}. 

%~/common/T_map/dtrap
%~/common/muppter/synthetic_dtrap/rot.pl ./synthetic_dtrap/genfig_maps.pl
%rsync -va fig_maps.pdf  ../fig_maps_input.pdf 
%bol10:30:56~/common/muppet/synthetic_dtrap$ rsync -va ~/common/T_map/dtrap/data/necplusultra/Tmap_tau1.fits . 

{\it Facilities:} \facility{ATCA, ALMA}.

\bibliography{/Users/simon/common/texinputs/merged.bib}

\end{document}